\documentclass[prb,preprint,amsmath,amssymb]{revtex4}
\begin{document}
\author{Kevin Leung,$^1$ Ida M.B. Nielsen,$^2$ Na Sai,$^3$ Craig Medforth,$^4$
and John A.~Shelnutt$^5$}
\affiliation{$^1$MS 1415, Sandia National Laboratories, Albuquerque,
NM 87185, USA}
\affiliation{$^2$MS 9158, Sandia National Laboratories, Livermore,
CA 94551, USA}
\affiliation{$^3$Department of Physics, University of Texas at Austin,
TX 78712, USA}
\affiliation{$^4$Department of Chemical and Nuclear Engineering,
University of New Mexico, NM 87131, USA}
\affiliation{$^5$MS 1349, Sandia National Laboratories, Albuquerque,
NM 87185, USA}
\date{\today}
\title{Cobalt-Porphyrin Catalyzed Electrochemical
Reduction of Carbon Dioxide in Water II: Mechanism from First
Principles}

\begin{abstract}

We apply first principles computational techniques to analyze the
two-electron, multi-step, electrochemical reduction of CO$_2$ to CO
in water using cobalt porphyrin as a catalyst.  Density Functional
Theory calculations with hybrid functionals and dielectric continuum
solvation are used to determine the steps at which electrons
are added.  This information is corroborated with {\it ab initio}
molecular dynamics simulations in an explicit aqueous environment
which reveal the critical role of water in stabilizing a key intermediate
formed by CO$_2$ bound to cobalt.  Using potential of mean force
calculations, the intermediate is found to spontaneously accept a proton
to form a carboxylate acid group at pH$<$9.0, and the subsequent cleavage of
a C-OH bond to form CO is exothermic and associated with a small 
free energy barrier.   These predictions suggest that the proposed
reaction mechanism is viable if electron transfer to the catalyst is
sufficiently fast.  The variation in cobalt ion charge and spin
states during bond breaking, DFT+U treatment of cobalt $3d$ orbitals,
and the need for computing electrochemical potentials are emphasized.

\end{abstract}

\maketitle

\input epsf

\section{Introduction}
 
CO$_2$ capture from flue gas and its conversion to useful products,
including fuel molecules, has emerged as an important paradigm for a
carbon-neutral economy.\cite{co2rev1,co2rev0,co2rev2,co2rev3}
At discussed in the preceeding paper of this series
(henceforth "paper I"),\cite{paper1} high ($\sim$70\%) yield of carbon monoxide
(CO) has been demonstrated in cobalt macrocycle-catalyzed electrochemical
reduction of carbon dioxide (CO$_2$) in water at applied voltage of about
-1.0~volt.\cite{furuya,ryba,sonoyama,magd2,magd1,ramirez1,ramirez2,chilean1,chilean2}  
The mechanisms of CO$_2$ reduction in non-aqueous solvents,
for which much more negative potentials are needed, have
been examined using a variety of methods.\cite{fujita02,grod00a,grod00b,fujita00c,fujita99,fujita98,fujita97,fujita95,fujita93a,fujita93b,fujita91b,fujita91a,fujita89,fujita09}

Co(I)P-catalyzed CO$_2$ reduction in water and protic solvents, which requires
a much less negative voltage for the onset of reaction than in organic
solvent,\cite{furuya,magd1,magd2,sonoyama,ramirez1,ramirez2,chilean1,chilean2,ryba}  
has received less fundamental studies.  The present
theoretical work focuses on the mechanism of this electrochemical
reaction in aqueous media.  As discussed in the preceeding paper in
this series\cite{paper1} (henceforth ``Paper I'') which examines the
structures, energetics, and charge states of reaction intermediates in detail,
the reaction likely takes place in the following logical sequence of steps:
\begin{eqnarray}
{\rm CoP} + {\rm CO}_2 &\rightarrow& {\rm CoPCO}_2 \, ; \label{eq1} \\
{\rm CoPCO}_2 + {\rm H}^+ &\rightarrow& {\rm CoPCOOH} \, ; \label{eq2} \\
{\rm CoPCOOH}  &\rightarrow& {\rm CoPCO} + {\rm OH}^- \, ; \label{eq3} \\
{\rm CoPCO} &\rightarrow& {\rm CoP} +  {\rm CO} \, . \label{eq4} 
\end{eqnarray}
``CoP'' will henceforth denote cobalt porphine, the simplest porphyrin
species, adopted in this work for ease of calculations.\cite{note98}  
Key questions to be addressed include: (1) why the
reaction, which involves protonation to form carboxylate acid motifs (COOH)
with p$K_{\rm a}$ typically on the order of 4.5, readily proceeds despite
the fact that pH$> 7$ in experiments;
(2) whether all steps are thermodynamically downhill; (3) whether
the free energy barriers of the intermediate steps are low enough
to be consistent with the observed reaction rate; and (4) at what
stages the two electrons are added.

In Eqs.~\ref{eq1}-\ref{eq4} above, we have
intentionally left out the charge states of the intermediates as yet
unassigned in experiments.  Since the two electrons can be added at any
step(s), numerous mechanisms are consistent with these equations.
The sequence of electron injection is governed by the redox potentials 
($\Phi_{\rm redox}$) of
the pertinent reaction intermediates relative to the applied voltage.
The half-cell potentials of various charge states of cobalt porphyrins
are known,\cite{reference_book} but not those for the CO- and CO$_2$-
bound complexes.

In this work, we use a combination of {\it ab initio} molecular dynamics
(AIMD)\cite{cpmd} techniques and Density Functional Theory (DFT) 
calculations with various exchange-correlation functionals and the
polarizable continuum model\cite{pcm}
(hereafter referred to as ``DFT+pcm'') to study 
Eqs.~\ref{eq1}-\ref{eq4}.  These methods inform and support each other.
The more economical DFT+pcm calculations approximate the aqueous
environment as a dielectric continuum,\cite{paper1} allowing a global
overview of the entire electrochemical reaction and a survey of the numerous
reaction intermediates.  Thus, DFT+pcm results reported in Paper~I are used to
extract $\Phi_{\rm redox}$ for all possible intermediates.  Redox potentials
consist of hydration free energy ($\Delta G_{\rm hyd}$) and ionization
potential/electron affinity contributions.  While $\Phi_{\rm redox}$ and
$\Delta G_{\rm hyd}$ have also been calculated with the AIMD method and explicit
treatment of the aqueous environment,\cite{ion,redox1,redox3,redox5}
these have so far been limited to monoatomic ions and molecules much
smaller than porphyrins comprising a minimum of 37 atoms.\cite{quasi}
Our B3LYP DFT+pcm redox potentials predict that CoPCOOH$^-$ is the key
intermediate.  To corroborate this $\Phi_{\rm redox}$ conclusion, we also
conduct more costly AIMD simulations with explicit water molecules.
The hydration structures of the reaction intermediates predicted therein
help explain the redox potential trends.

Next, we perform AIMD calculations on the individual reactions identified
as the key steps in DFT+pcm calculations.  Although it is possible that
the electron addition steps of the two-electron reduction of
CO$_2$ are rate-limiting,\cite{chilean1,chilean2}
modeling the electron transfer rate via Marcus theory approach\cite{marcus}
may not answer the most interesting scientific questions.  This is
because the electron injection rate almost certainly depends on engineering
aspects such as the electrical contact between the gas diffusion
electrode and the polymerized catalyst.  Besides, it is necessary to
demonstrate the viability of other steps in which electron transfer is
not involved.  We omit the electrode, focus on cobalt porphine molecule
dispersed in water, and use AIMD
to study two reactions involving [Co(II)PCOOH]$^-$: deprotonation (the
reverse of Eq.~\ref{eq2}) and the cleavage of the C-OH bond (Eq.~\ref{eq3}).
The protonation reaction proves critical to the
efficient removal of one of the CO$_2$ oxygen atoms and reduction
of the carbon atom from the +4 to the +2 formal charge state.  The
p$K_{\rm a}$ of [Co(II)PCOOH]$^-$ will be determined using AIMD potential
of mean force (PMF) in a manner previously applied to silanol groups on
silica surfaces.\cite{silica}  As for the free energy change and barrier
associated with the breaking of the C-OH bond (Eq.~\ref{eq3}) prior to
releasing CO (Eq.~\ref{eq4}), there is as yet no experimental reaction rate
for direct comparison.  The turnover rate per catalytic active site is obscured
by the undetermined proportion of active CoP molecules actually participating in
the reactions.  But our PMF with an approximate reaction coordinate
yields a barrier which suggests that CO gas evolution should proceed
readily if electron transfer from the electrode is fast.
It is known experimentally that Co(II)P binds weakly to CO;\cite{copco,mu1989}
our gas phase calculations also shows that the Co(I)P-CO complex is weakly
bound.  Therefore once the C-OH bond is severed to form a OH$^-$, the
subsequent steps (Eq.~\ref{eq4}) should be exothermic, fast, and
non-rate-determining, and they do not require further theoretical studies.  

This work provides fundamental understanding specific to the CO$_2$ reduction
mechanism in water using macromolecule catalysts, highlights the
critical role played by the protic solvent water in lowering the voltage
needed for the reaction, and may shed light on ways to further
improve and modify the cobalt porphyrin catalyst.  It is also of general
interest to the fledging field of computational electrochemistry.  
The multistep nature of the reaction emphasizes the importance of accurate
calculation of redox potentials.\cite{brookhaven,b3lyp_redox,truhlar04,redox2}
The protonation reaction (Eq.~\ref{eq2}) is accompanied with a change
in the Co ion charge state, which can be considered a form of coupled
proton/electron process.\cite{brookhaven,hammes-schiffer}  We show that
this can induce hysteresis in AIMD simulations.  Finally, this study in
explicit water, comprising AIMD trajectories exceeding 200~ps in duration,
is greatly facilitated by the use of an empirical DFT+U method\cite{dftu}
that can treat the localized $3d$-orbitals of the Co ion accurately without
resorting to more costly theoretical methods such as hybrid functionals.

This paper is organized as follows:  The theoretical methods used are
discussed in Sec.~2.  Section~3 describes the redox potential
predictions which determine the electron addition steps, and then
focuses on the deprotonation and C-O bond cleavage reactions of the
key intermediate [Co(II)PCOOH]$^-$.  Section~4 compares the method used
in the present work to our previous CO$_2$-related simulations;\cite{co2} it
further looks to the future and briefly discusses new computational techniques
that may facilitate the modeling of demanding electrochemical processes.
The accuracy of electronic structure calculations which underlie our
mechanistic predictions will also be addressed.\cite{truhlar_rev}  Section~5
concludes and summarizes the paper.  An appendix discusses minor
hysteresis issues encountered in one part of the calculations.

\section{Method}
\label{method}

All B3LYP or PBE plus dielectric continuum calculations apply the
Gaussian suite of programs
version g03.\cite{gauss}  All DFT+U calculations
use the VASP code.\cite{vasp,vasp1}  The Supporting Information document
(SI) Sec.~S1 provides a brief comparison between these packages.

DFT plus dielectric continuum (DFT+pcm) calculations apply 
the Becke-3-parameter-Lee-Yang-Parr
(B3LYP)\cite{becke1993,lee1988} or the Perdew-Burke-Ernzerhof
(PBE)\cite{pbe} functional,  and the PCM dielectric continuum
model.\cite{pcm}  Other details are described in Paper~I.\cite{paper1}
Zero-point energy (ZPE) contributions to Eqs.~\ref{eq2} and~\ref{eq3} are
estimated at T=0~K, also using Gaussian\cite{gauss} and the 6-31+G$^*$
basis set.  The H$_2$O-OH$^-$ complex is assumed to be the H$^+$
accepting or OH$^-$ containing species in ZPE calculations.

The redox potential of the A$^{(n+1)-}$/A$^{n-}$ couple is the electron
affinity (EA) of A$^{n-}$ plus the difference in hydration free energies
($\Delta G_{\rm hyd}$ difference between A$^{(n+1)-}$ and A$^{n-}$).
Both EA and $\Delta G_{\rm hyd}$ are readily obtained from DFT+pcm total
free energies that include
zero point energy corrections and finite temperature contributions.
Where available, DFT+U $\Phi_{\rm redox}$ are estimated by substituting
ground state DFT+U energies for B3LYP ones but retaining B3LYP
dielectric hydration and ZPE information.\cite{note9} All
reported $\Phi_{\rm redox}$ are referenced to the accepted value of
4.44~volt for the standard hydrogen electrode half-cell potential.  
The effect of the choice of DFT functionals on $\Phi_{\rm redox}$
will be addressed in Sec.~\ref{discussion}.

Spin-polarized AIMD simulations apply Vienna ab-initio simulation
program (VASP),\cite{vasp,vasp1}
the PBE functional, $\Gamma$-point Brillouin zone sampling, 400~eV planewave
energy cutoff, deuterium masses for all protons to allow Born-Oppenheimer
dynamics time steps of 0.25~fs, a 10$^{-6}$~eV energy convergence criterion,
and T=425~K NVT conditions using a Nose thermostat.  At T=400~K,
the PBE functional yields average pure water structure consistent with
experimentally observed water $g(r)$ at T=300~K.\cite{water_pbe}  Since
the porphine ring exhibits significant ruffling fluctuations that explore
a large configurational space, we raise the temperature by an extra 25~K
to promote better sampling statistics.  AIMD simulations apply
13.64~\AA$\times$13.64~\AA$\times$ 13.64~\AA\, simulation cells
which contain a CoPCOOH complex and 71~H$_2$O molecules.\cite{glycine}

Semi-local functionals such as PBE are generally inadequate for treating first
row transition metal complexes like cobalt,\cite{reiher2001,ghosh2005}
although they are more successful with transition metal dimers.\cite{tmdimer}
In general, transition metal complexes have been a challenge to DFT
methods.\cite{truhlar_rev}
We apply the following reaction to benchmark the prediction energetics:
\begin{equation}
{\rm [Co(III)P]}^+ + {\rm CO} \rightarrow {\rm [Co(III)P-CO]}^+. \label{eq7}
\end{equation}
While Co(II)P itself seems well represented by DFT
functionals,\cite{rovira1,jaworska2007}
Paper~I has shown that the widely used PBE and B3LYP functionals predict
Eq.~\ref{eq7} binding energies which differ from the experimental
value\cite{mu1989} by about $0.5$~eV in opposite directions.  To deal
with this problem, we augment the PBE functional with DFT+U\cite{dftu}
applied to the partially occupied $3d$-orbitals of the Co ion.  With
a judicial choice of the $U$ parameter, this approach has been shown to
give accurate predictions for organometallic compounds, although agreement
between theory and experiments has not been
universal.\cite{porph,mnp_h2o,kulik,sit,sit1,cocc,pooja1,pooja2} Using VASP
PAW pseudopotentials,\cite{vasp1} setting $U$=2.5~eV yields a binding energy
for Eq.~\ref{eq7} that agrees with experimental data reported for
the Co(III)TPP-CO complex.\cite{paper1,mu1989,note9}  At T=0~K, these two
complexes exhibit very similar VASP/PBE binding energies of 1.319 and
1.351~eV respectively; they differ by only 0.74~kcal/mol, suggesting that the
experimental Co(III)TPP-CO binding free energy is a good metric for
benchmarking the theoretical Co(III)P-CO predictions.  As neither CO$_2$ nor
CO strongly binds to cobalt porphyrins at most accessible Co charge states,
Eq.~\ref{eq7} is the only binding constant available in the experimental
literature as a benchmark.  While this value of $U$ may not be optimal for
all Co charge states, referencing to Eq.~\ref{eq7} appears the most
justifiable empirical route.  The accuracy of AIMD simulations depend
on the DFT+U method used, as will be discussed in Sec.~\ref{discussion}.

As reported in Paper I,~\cite{paper1} the optimal spin states of all
gas phase B3LYP and PBE CoP complexes are low spin except [Co(I)PCOOH]$^{2-}$
and the unligated [Co(III)P]$^+$, both of which are triplets.  DFT+U
calculations at T=0~K yield the same optimal spin states.

Equation~\ref{eq2} involves calculating the p$K_{\rm a}$ of
[Co(II)PCOOH]$^-$.  p$K_{\rm a} = -\log_{10}$ $\exp(-\beta \Delta G^{(0)})$
has been successfully computed for molecules and surfaces in liquid water
using the AIMD technique.\cite{sprik,klein,parrin1,sprik_new_proton,silica}
Here $\beta$ is $1/k_{\rm B}T$, and $\Delta G^{(0)}$ is the standard state
deprotonation free energy,
\begin{equation}
\Delta G^{(0)} = -k_{\rm B}T {\rm ln} \bigg\{ C_0
        \int_0^{R_{\rm cut}} dR \, A(R) \,
        \exp[-\beta W(R)] \bigg\} \, . \label{eq8}
\end{equation}
$C_0$ denotes 1.0~M concentration, $R$ is the reaction coordinate,
$A(R)$ is a phase space factor to be discussed in Sec.~\ref{results},
$R_{\rm cut}$ is the cutoff distance delimiting the reaction and product
valleys in the free energy landscape, and $W(R)$ is the potential of
mean force (PMF), referenced such that $W(R)=0$ as $R \rightarrow \infty$.
$R_{\rm cut}$ is taken as the onset of the plateau where
$W(R) \rightarrow 0$.  The umbrella sampling technique\cite{book1} and
a 4-atom reaction coordinate
\begin{equation}
R=R_1-R_2-R_3 \label{R_def}
\end{equation}
are applied to compute $W(R)$.  Here $R_1$ is the distance between the
COOH acid proton and the oxygen atom on the designated proton-accepting
H$_2$O, while $R_2$ and $R_3$ are the distances between this O~atom and
the two protons originally on the designated H$_2$O molecule,
respectively.\cite{silica}  As $R_2$ and $R_3$ are about 1.0~\AA\, for
intact O-H covalent bonds, it can be readily inferred that $R \sim -1.0$~\AA\, 
is consistent with deprotonation and CO$^-$/H$_3$O$^+$ contact ion pair
formation while $R > -0.4$~\AA\, indicates an intact CO-H bond.\cite{silica}
Designating a special H$_2$O molecule can be done without loss of
generality because all water molecules are interchangeable and only one
is at any time is close enough to the acid proton to be considered
a potential proton acceptor.  Harmonic potentials of the form
\begin{equation}
U(R)=B (R-R_o)^2 , \label{uofr}
\end{equation}
with $B$ values between 2 to 4~eV/\AA$^2$ are applied to 7 umbrella windows
with $R_o$ spanning the range of $R$ to be sampled.  We reference the
p$K_{\rm a}$ of [CoPCOOH]$^-$ relative to the free energy of
water-autoionization,\cite{klein} computed using the same reaction
coordinate and elevated temperature and assumed to exhibit p$K_{\rm w}$=14.  
As an validation test, our AIMD p$K_{\rm a}$ methodology has
been applied to formic acid in water, yielding an acidity constant within
a fraction of a pH unit of the experimental value (SI Sec.~S2).

A ``reflecting boundary condition'' potential $V(R_{\rm OH})$ sets an
approximate 1.2~\AA\, distance of closest approach between water protons
and the hydroxyl oxygen atom.  It preserves the identity of the deprotonated
-COOH (or H$_2$O in the case of water auto-ionization) by preventing proton
transfer via the Grotthuss mechanism.  
$V(R_{\rm OH})=B (R_{\rm OH}-R_1)^4$, where $B$=200~eV/\AA$^4$ and
$R_1$=1.3~\AA, is imposed whenever $R_{\rm OH} < R_1$.  $R_{\rm OH}$
is the distance between the hydroxyl (COH) oxygen and all H$^+$ other than  
the original COOH proton.  Related boundary potentials have been
applied to AIMD simulations of other chemical reactons.\cite{klein1,co2}
$V(R_{\rm OH})$ is only necessary in the deprotonation umbrella sampling window
with the most negative $R$.  As this leftmost umbrella sampling window
exhibits a $W(R)$ variation of only $\sim 0.6$~kcal/mol
(see Sec.~\ref{results}), the effect of $V(R_{\rm OH})$ should be small ---
much less than 0.6~kcal/mol.  

\begin{figure}
\centerline{(a) \hbox{\epsfxsize=1.5in \epsfbox{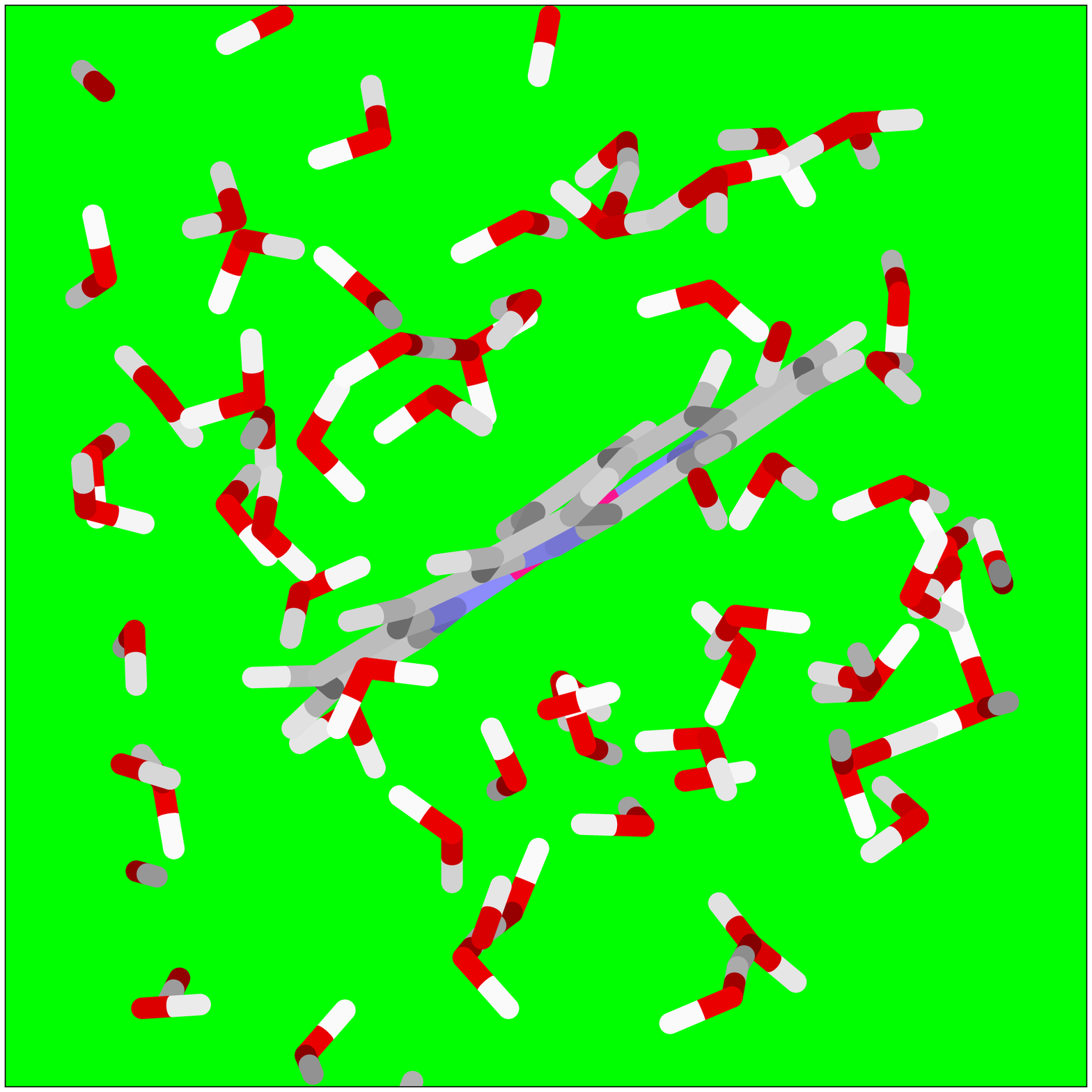}}
            \hbox{\epsfxsize=1.50in \epsfbox{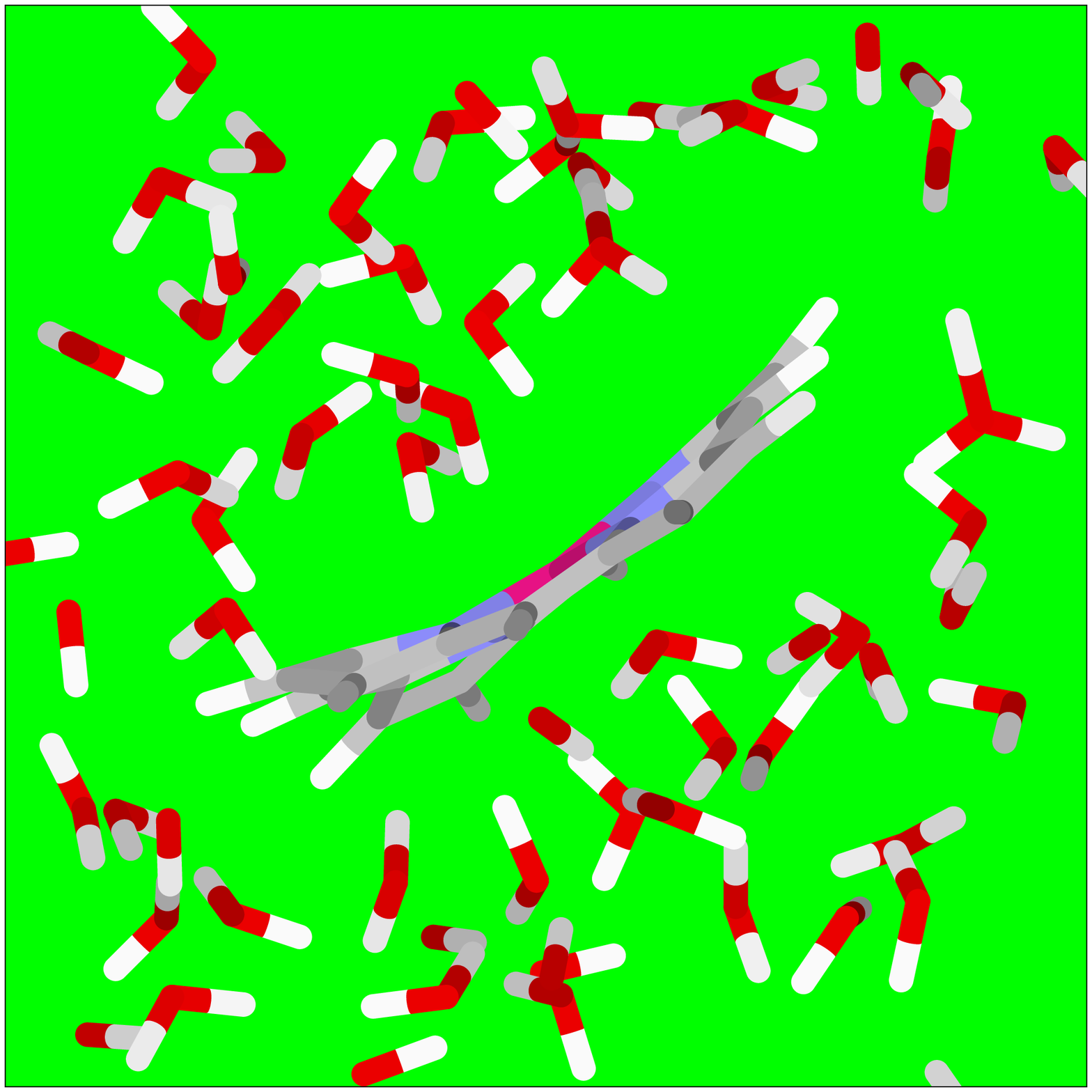}} (b)}
\centerline{(c) \hbox{\epsfxsize=1.50in \epsfbox{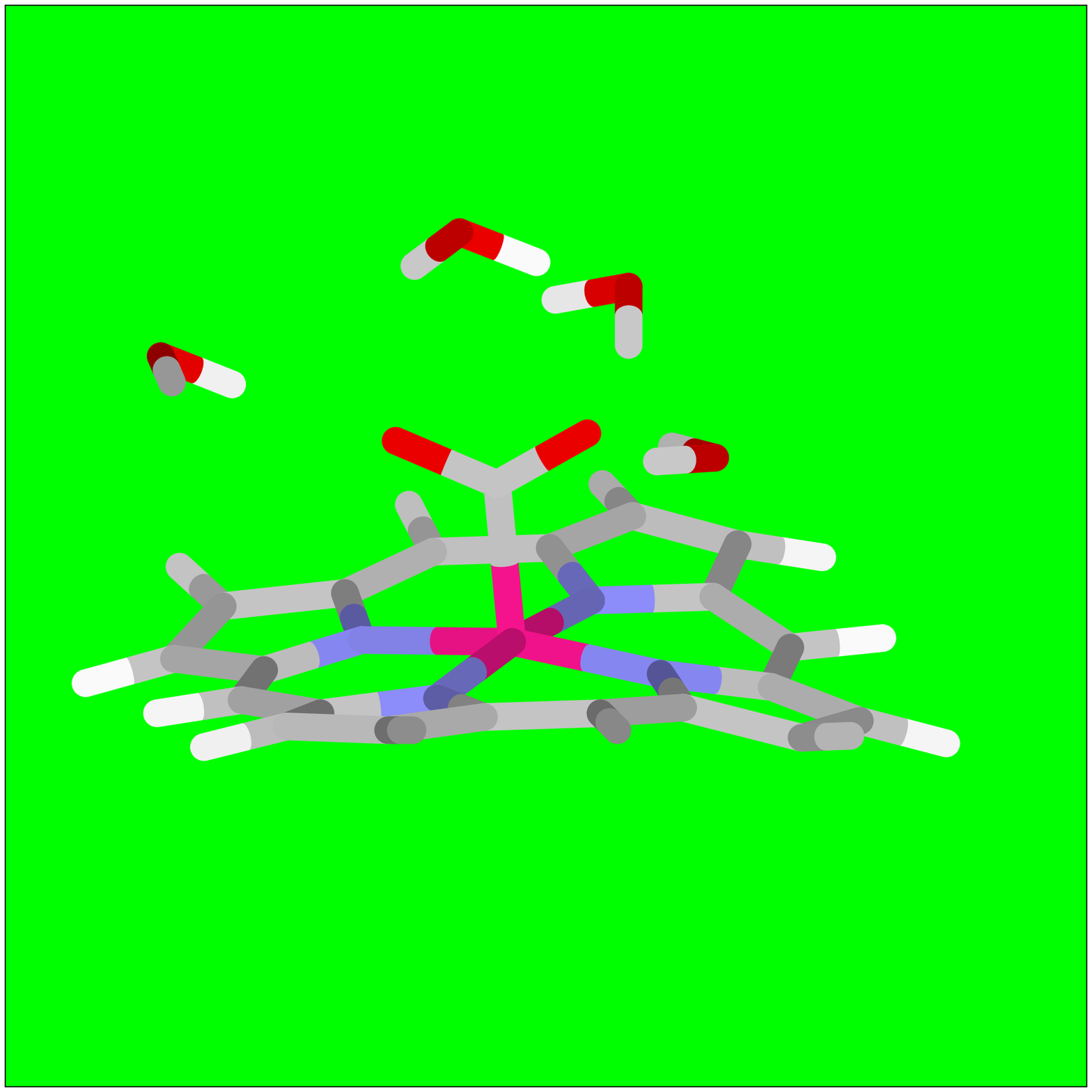}}
            \hbox{\epsfxsize=1.50in \epsfbox{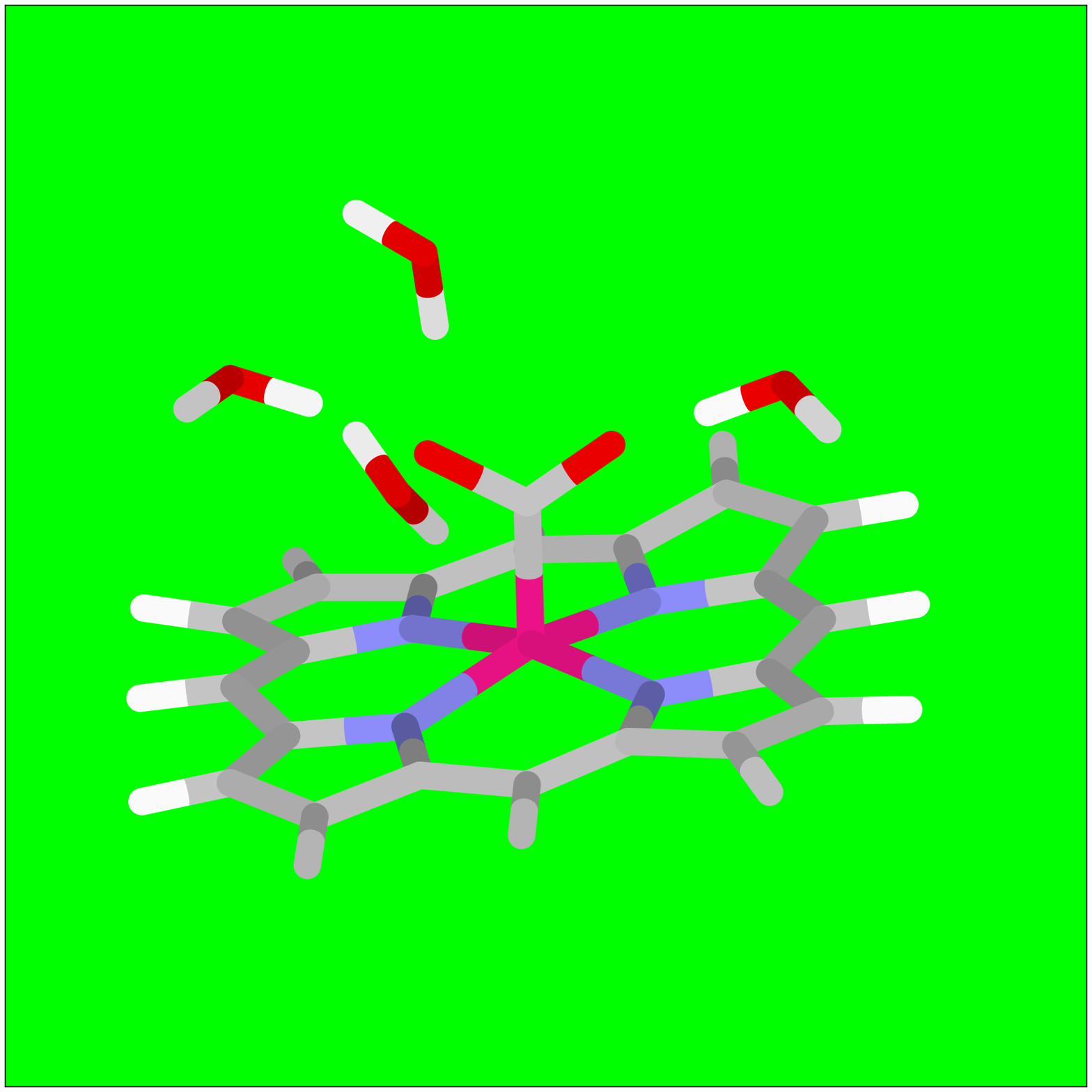}} (d)}
\centerline{(e) \hbox{\epsfxsize=1.50in \epsfbox{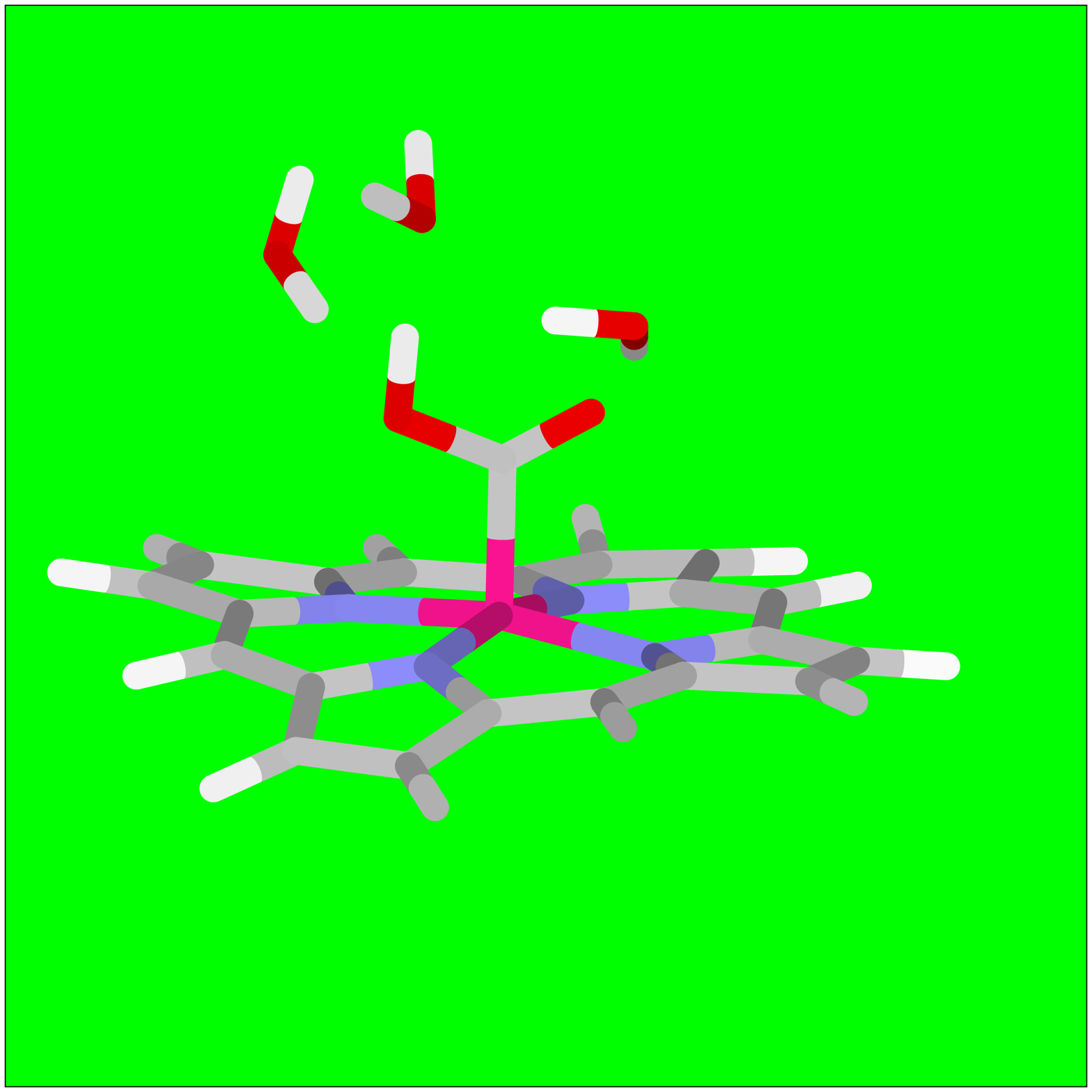}}
            \hbox{\epsfxsize=1.50in \epsfbox{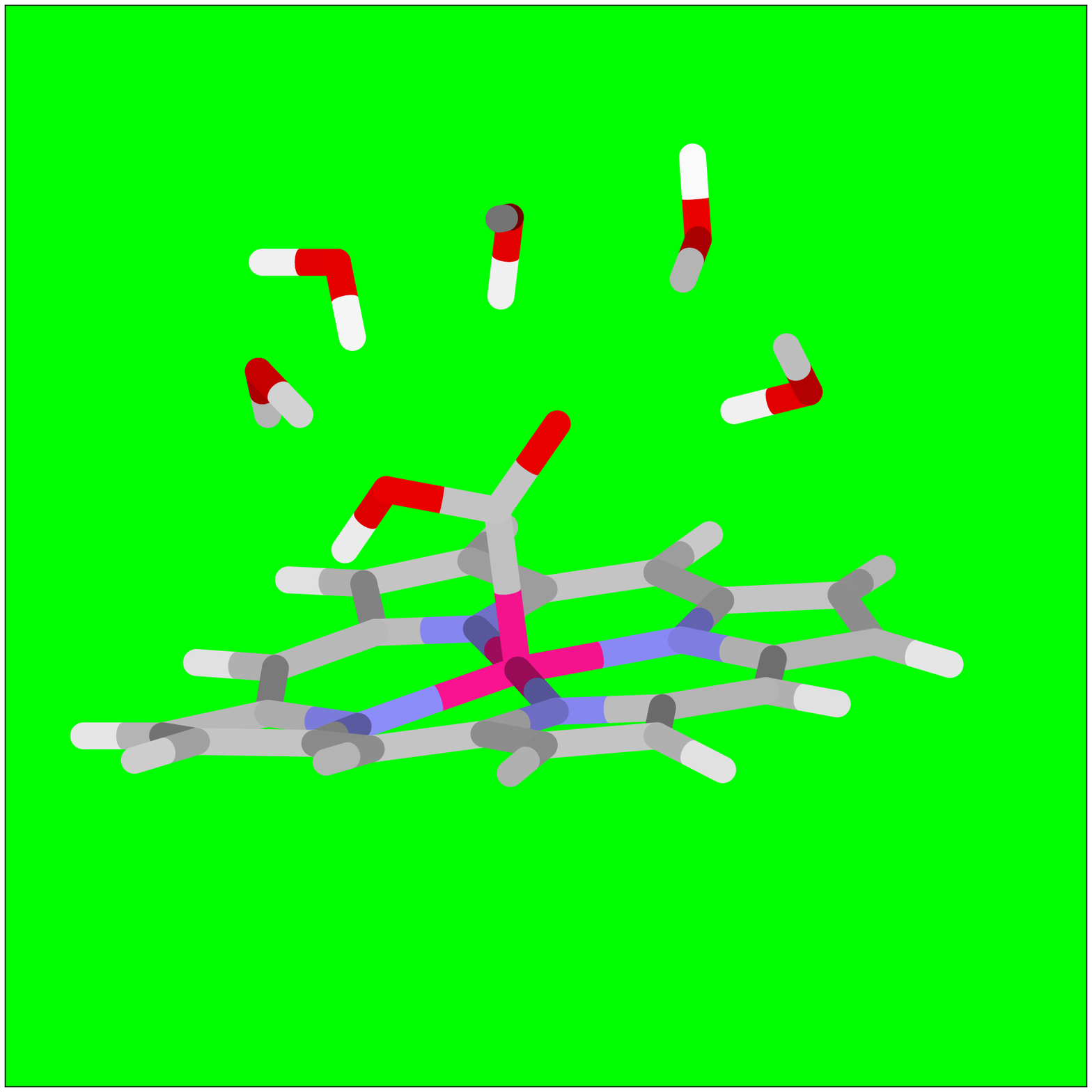}} (f)}
\caption[]
{\label{fig1} \noindent
Snapshot of (a) [Co(I)P]$^-$; (b) [Co(I)P]$^{2-}$
(see footnote~\onlinecite{note_cop} for description of charge states;
(c) [Co(I)PCO$_2$]$^-$; 
(d) [Co(I)PCO$_2$]$^{2-}$; (e) Co(II)PCOOH; (f) [Co(II)PCOOH]$^-$; in water.
Panels (a)-(b) show the absence of hydrogen bonding between the negatively
charged CoP and water.
In (c)-(f), most H$_2$O molecules are omitted; only those forming
hydrogen bonds with the CoP structure are shown.  Panels (e) and (f)
depict the C-OH group in the {\it trans} (exo) and {\it cis} (endo)
configuration, respectively.  All snapshots are taken after 1-2 ps of
short AIMD runs.  Pink: Co; grey: C; red: O; white: H.
}
\end{figure}

For the C-OH cleavage reaction (Eq.~\ref{eq3}), 
the reaction coordinate is taken to be
$R_{\rm C-O}$, the distance between the COOH carbon and the
hydroxyl oxygen atom.  
Harmonic potentials of the form Eq.~\ref{uofr} are
applied to 10 umbrella sampling windows, but with $R_{\rm C-O}$
replacing the 4-atom coordinate $R$ in this case.  The phase space
factor $A(R)$ in the free energy expression analogous to Eq.~\ref{eq8} 
becomes $4\pi R_{\rm C-O}^2$.

Dealing with slow, diffusive degrees of freedom is a significant challenge
in AIMD calculations of bond-breaking free energies.\cite{co2,klein1}  
During the stretching/cleavage of the C-OH bond, rotational phase factors of
$k_B T \log [4 \pi (R_{\rm C-O})^2]$ should naturally emerge in
well-converged evaluations of Eq.~\ref{eq3}.  However, rotation of 
the nascent product OH$^-$ ion about the carbonyl carbon atom in water is
too slow on AIMD time scales to accurately reproduce this rotational
entropy.  To give a better converged $W(R)$, the $x$- and $y$-coordinates
of the carbonyl C and hydroxyl O atoms are kept identical
and fixed while their $z$ coordinates are allowed to vary.  No other
atom is frozen in AIMD simulations.  The rotational entropy is restored
by multiplying $4\pi R_{\rm C-O}^2$ to the $R_{\rm C-O}$ probability
distribution function at each
$R_{\rm C-O}$.\cite{klein1,co2,note99}  The SI, Sec.~S3, shows that
this constraint has modest effect on the sampling of the Co-C-O angular
distribution despite the bulkiness of the CoP group.

The deprotonation free energy change in Eq.~\ref{eq2} is a state
function which should not depend on the reaction coordinate chosen
provided that equilibrium sampling is achieved.  Using the one-dimensional
coordinate $R$, however, we observe some hysteresis due to the picosecond
time scale relaxation of the charge state of the cobalt ion as the extent
of deprotonation varies.  For Eq.~\ref{eq3}, we are
interested in the free energy barrier in addition to the
free energy change.  Umbrella sampling yields a free energy
barrier estimate which depends on and is generally underestimated
by any chosen reaction coordinate because some
trajectories with forward velocity can recross the ``transition state''
point and do not proceed to product formation.\cite{book1,chandler78}
To assess the validity of the computed barrier, we perform transmission
coefficient ($\kappa$) calculations,\cite{book1,chandler78} to be
discussed in more detail below.  The transition path sampling
method\cite{chandler} is the rigorous approach to compute free energy
barriers; although more costly, it will be considered in the future.
Multi-dimensional metadynamics,\cite{meta1,meta2,meta3}
new deprotonation coordinates,\cite{sprik_new_proton} the self-consistent
DFT+U method,\cite{cocc} and new DFT functionals\cite{truhlar_tm1}
may also benefit future AIMD-based electrochemical
calculations, and will be mentioned in Sec.~\ref{discussion}.

The amount of water in the simulation cell is determined using grand
canonical Monte Carlo simulations at constant water chemical potential,
the Towhee Monte Carlo code,\cite{towhee} and the SPC/E water model.\cite{spce} 
The cell size is identical to that used in AIMD simulations (13.64~\AA$^3$).
The temperature is set at T=300~K because the SPC/E model, unlike DFT/PBE,
yields reasonable water structure at room temperature.  One charge-neutral
Co(III)PCOOH is placed frozen in its DFT-optimized configuration in
the simulation cell.  The CoP and COOH Lennard-Jones force field parameters
are approximated with those of Mn(II)P\cite{shelnutt_para} 
and the formate anion, respectively.  The atomic
partial charges are assigned using Mulliken charge analysis of a gas phase
B3LYP/6-311+G(d,p) Gaussian calculation.  $4\times 10^8$ Monte Carlo
moves are attempted, 30\% of them being water insertion/deletions.
The most probable number of water molecules in the simulation cell is
determined to be 71, and this is the water content used in AIMD
simulations.  A net $-|e|$ charge and a neutralizing
background are then imposed on the final CoPCOOH/H$_2$O configuration
from the Monte Carlo run, and AIMD simulations are initiated.  We choose a
few sampling windows along the reaction coordinates as seed windows.
With the appropriate umbrella sampling potentials of the selected windows
turned on, AIMD pre-equilibration is conducted for 2~ps at T=500~K.  Maximally
localized Wannier function analyses\cite{wannier} confirm that the extra
electron resides on CoPCOOH.  Then the system is further equilibrated at the
target temperature T=425~K for another 2~ps before statistics are
collected.  The starting configurations for all other sampling windows are 
taken successively from snapshots of adjacent windows a few picosecond
into their AIMD trajectories.

\begin{table}\centering
\begin{tabular}{||l|l|l|l|l||} \hline
reduced & oxidized & B3LYP & PBE & DFT+U \\ \hline
Co(I)P$^-$ & Co(II)P & -1.71 & -0.89 & -1.46$^*$ \\
Co(I)P$^{2-}$ & Co(I)P$^-$ & -2.11 & -2.31 & NA \\
Co(I)PCO$_2$$^{2-}$ & Co(I)PCO$_2$$^-$ & -1.82 & -1.78$^*$ & -1.68$^*$ \\
Co(II)PCOOH$^-$ & Co(III)PCOOH & -1.17 & -2.14 & -1.07$^*$ \\
Co(II)PCOOH$^{2-}$ & Co(II)PCOOH$^-$ & -2.17 & -2.08$^*$ & NA \\ \hline
\end{tabular}
\caption[]
{\label{table1} \noindent
Redox potential of various species using three functionals/computational
methods, in volts.  ZPE are computed using the B3LYP functional; asterisks
indicate that dielectric solvation contributions also come from B3LYP.
Convergence cannot be achieved when applying DFT+U 
to some doubly negatively charged systems in the gas phase.  The
electron affinities of these species are listed in the SI, Sec.~S4.
DFT+U calculations are performed using VASP; B3LYP and PBE calculations
apply the Gaussian suite of codes.}
\end{table}

The statistical uncertainty in each sampling window is estimated by splitting
the trajectory into four equal parts, calculating the standard deviation
for $W(R_2)-W(R_1)$, where $R_1$ and $R_2$ are the boundary values in
the window, and then dividing by $\sqrt{4}$ to yield an approximate 
error bar for the entire trajectory.  The overall uncertainty convolves the
standard deviations in all windows.  The statistics are generated with
AIMD trajectories of at least 10~ps in duration; in a few windows where the 
uncertainties are large, 15-20~ps AIMD runs are conducted.

The maximally localized Wannier function analysis\cite{wannier} is used to 
determine the charge/spin state of the cobalt ion in the reactants,
productions, and reaction intermediates in liquid water.  As discussed
in the SI Sec.~S1, the Mulliken charge decomposition technique, 
popular in the quantum chemistry community, has not been
implemented in the planewave basis VASP code.  Alternative charge
decomposition methods such as simply integrating the
charge/spin densities within some radius around the transition metal
ion tend to yield ambiguous results.\cite{zunger}  Hence the Wannier
approach appears the most useful method in our condensed phase, periodically
replicated simulation cell setting.

\section{Results}
\label{results}

\subsection{Redox Potentials, Hydration Structures, and Electronic Structures}

The redox potential values are listed in Table~\ref{table1}.  
To anticipate the conclusions, we find that the predicted B3LYP
DFT+pcm absolute redox potentials are not
in good agreement with experimental values.  However, these potentials
referenced to one redox couple yield qualitative results consistent with
insights gained from AIMD simulations, and they allow us to assign the
charge state of the key intermediate species.

Thus, at first glance, B3LYP appears to predict a Co(I)P/Co(II)P redox couple
value which disagrees with the average experimental value of $-0.67$~volt
($-0.5$ to $-0.84$, depending on the porphyrin ring substituent and
solvent\cite{reference_book}).   The redox potential also strongly
depends on the DFT functional used, in contrast to the findings of
Ref.~\onlinecite{jaworska2007} for the Co(II)P-NO/[Co(III)P-NO]$^+$
couple.  DFT+U and B3LYP redox potentials track each other while the PBE
functional yields substantially different values.  Systematic errors in
hybrid DFT plus dielectric continuum redox potentials have been reported
in the literature,\cite{b3lyp_redox,redox2}
particularly those associated with the B3LYP functional.\cite{b3lyp_redox}
While part of the discrepancy in Table~\ref{table1} may be due to DFT
inaccuracies, the DFT+U redox potential, fitted to Eq.~\ref{eq7}, is also
off by 0.8~volt.  Thus uncertainties arising from the use of the dielectric
approximation used to calculate $\Delta G_{\rm hyd}$ as well as the lack of
ring substituents in our calculations may also be responsible.  Although
PBE appears to yield a $\Phi_{\rm redox}$ for the [Co(I)P]$^-$/Co(II)P
couple close to the experimental value, it performs worse for the
benchmark reaction Eq.~\ref{eq7} than DFT+U.  The result for that
equation is deemed more reliable because the pertinent experiments
were performed in aprotic solvents which interact weakly with the
reactants.\cite{mu1989} 

\begin{figure}
\centerline{\hbox{\epsfxsize=3.2in \epsfbox{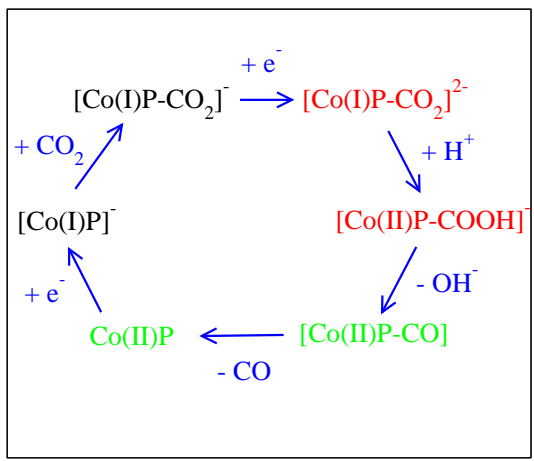}}}
\caption[]
{\label{fig2} \noindent
Mechanism of CO$_2$ reduction with electron addition deduced from hybrid
DFT plus dielectric continuum redox potential calculations.  Red denotes
key intermediates; green species should undergo fast reactions.
}
\end{figure}

To proceed, we focus on relative B3LYP redox potential values
and assume that all electrochemical measurements occur near the
B3LYP [Co(I)P]$^-$/Co(II)P voltage.  As discussed in the Introduction,
this corresponds to the experimental condition where the onset of
CO gas evolution is within a few tenths of a volt of the Co(II)P
reduction potential.  The B3LYP redox
potentials of other species relative to [Co(I)P]$^-$/Co(II)P ($-1.71$~volt)
then determine whether an additional electron has been incorporated at that
step.  A similar approach has been used in Ref.~\onlinecite{b3lyp_redox},
where it has also been suggested that relative B3LYP redox potentials
are much more reliable than absolute values.
Thus the B3LYP functional predicts that Co(III)PCOOH is reduced
to [Co(II)PCOOH]$^-$, while [Co(I)P]$^{-}$
and [Co(II)PCOOH]$^-$ are not reduced to [Co(I)P]$^{2-}$ and
[Co(II)PCOOH]$^{2-}$, respectively, because their required voltages are
much more negative than $-1.71$~volt.  The experimental reduction potentials
for [Co(I)P]$^-$ in aprotic solvents are indeed -0.51 to -1.2~volt more
negative than that for Co(II)P.\cite{reference_book}  However, unlike
[Co(I)P]$^-$ itself, [Co(I)PCO$_2$]$^-$ is already reduced
to [Co(I)PCO$_2$]$^{2-}$ at the [Co(I)P]$^-$/Co(II)P voltage
with a mere additional -0.11~volt.  In other words, it is much easier
to add an electron to [Co(I)PCO$_2$]$^-$ in water than [Co(I)P]$^-$.

These trends can be explained with DFT+U/AIMD simulations
in explicit water.  The snapshots in Fig.~\ref{fig1} are generated
using short, 1-2~ps AIMD trajectories, which are sufficient to
yield qualitative, well-equilibrated hydration structures if not highly
precise average hydration numbers.  Despite their net negative charges,
both [Co(I)P]$^-$ and [Co(I)P]$^{2-}$ are effectively hydrophobic plates
which do not form hydrogen bonds with water molecules;\cite{note_cop} no
water protons are observed within 2.5~\AA, a typical hydrogen bonding
cutoff distance, of the porphine N~and Co~atoms (Figs.~\ref{fig1}a-b).  
This is in contrast to the charge neutral and positively charged
Mn(II)P and Mn(III)P, where 1 or 2 H$_2$O molecule strongly
coordinates to the Mn site.\cite{mnp_h2o}
CO$_2$, a famously inert molecule, also fails to hydrogen-bond
with water.\cite{co2}  However, when they combine to form
[Co(I)PCO$_2$]$^-$ and [Co(I)PCO$_2$]$^{2-}$, the resulting
complexes form 4 to 5 hydrogen bonds with water through the partially
negatively charged O atoms on the CO$_2$, which now adopts a bent
geometry like a carbonate\cite{co2,kalinichev,rode} or a carboxylate
anion\cite{choo,daub} (Figs.~\ref{fig1}c-d).  The enhanced
interaction with water evidently facilitates the accommodation of an
extra electron.  This finding can be significant not just for electrochemical
reduction of CO$_2$, but also for CO$_2$ capture in general.\cite{tossell}
The B3LYP DFT+pcm results reflect this CO$_2^{n-}$ polarization information
despite the fact that water is treated implicitly there.

We also consider the hydration structures of Co(III)PCOOH and
[Co(II)PCOOH]$^-$ (Figs.~\ref{fig1}e-f).  The C-OH {\it trans} (exo)
and {\it cis} (endo) configurations are almost iso-energetic in gas phase
Co(III)PCOOH, within 1.06~kcal/mol of each other.
In contrast, [CoP(II)PCOOH]$^-$ forms a strong intramolecular hydrogen
bond between the COOH acid proton and one of the nitrogen atoms on the
porphine ring when the proton is in the {\it cis} position (Fig.~\ref{fig1}f).
This COOH proton cannot form hydrogen bond with other water molecules.
In the gas phase, this structure is
4.93~kcal/mol more stable than the {\it trans} configuration where the
OH points outwards (Fig.~\ref{fig1}e).  This feature will play a prominent
role in acid-base reactions (Eq.~\ref{eq2}).  Intramolecular hydrogen bonds
have also been suggested to facilitate CO$_2$ binding and chemical reduction
in the literature.\cite{co2rev0,co2rev3,fujita93b}

The calculated redox potentials (Table~\ref{table1}) suggest an overall
mechanism shown in Fig.~\ref{fig2}.   The CO$_2$ adsorption and the first
electron insertion steps are likely simultaneous and cooperative.  This
is because the B3LYP functional predicts that the following equation,
\begin{equation}
{\rm [Co(I)P]}^{2-} + {\rm CO}_2 \rightarrow {\rm [CoP(I)PCO_2]}^{2-} ,
\end{equation}
exhibits a very significant free energy gain of 27~kcal/mol in water
(treated as a dielectric continuum).\cite{paper1}  However, [Co(I)P]$^-$
should not be readily reduced to [Co(I)P]$^{2-}$ at or slightly below the
[Co(I)P]$^-$/Co(II)P half cell voltage, and CO$_2$ is not strongly bound
to [Co(I)P]$^-$.  This suggests that the CO$_2$ may be thought of as
part of the solvent, and the electron transfer to the cobalt complex as
a solvent/CO$_2$ fluctuation-mediated process akin to the Marcus theory
picture.\cite{marcus}  
As mentioned in the introduction, the electron transfer rate may depend
on the electrical contact between the catalyst and the gas-diffusion electrode,
and is not the main subject of this study.  The protonation and C-OH
cleavage steps in Fig.~\ref{fig2} are at the heart of the catalytic
function, and must be fast and spontaneous; understanding of the
scientific principles involved can lead to improved catalysts and
reaction conditions.  In the next subsections, we consider these steps
in detail using the AIMD method.  Co(II)P is known to be weakly bound
to CO.\cite{copco}  Although experimental data is not available,
B3LYP/6-31+G$^*$ calculations indicate that the gas phase Co(I)P-CO binding
energy is only 1.76~kcal/mol at T=0~K.  Once C-OH cleavage is achieved,
the rest of the reaction (Eq.~\ref{eq4}) should proceed rapidly.

\begin{figure}
\centerline{(a) \hbox{\epsfxsize=1.50in \epsfbox{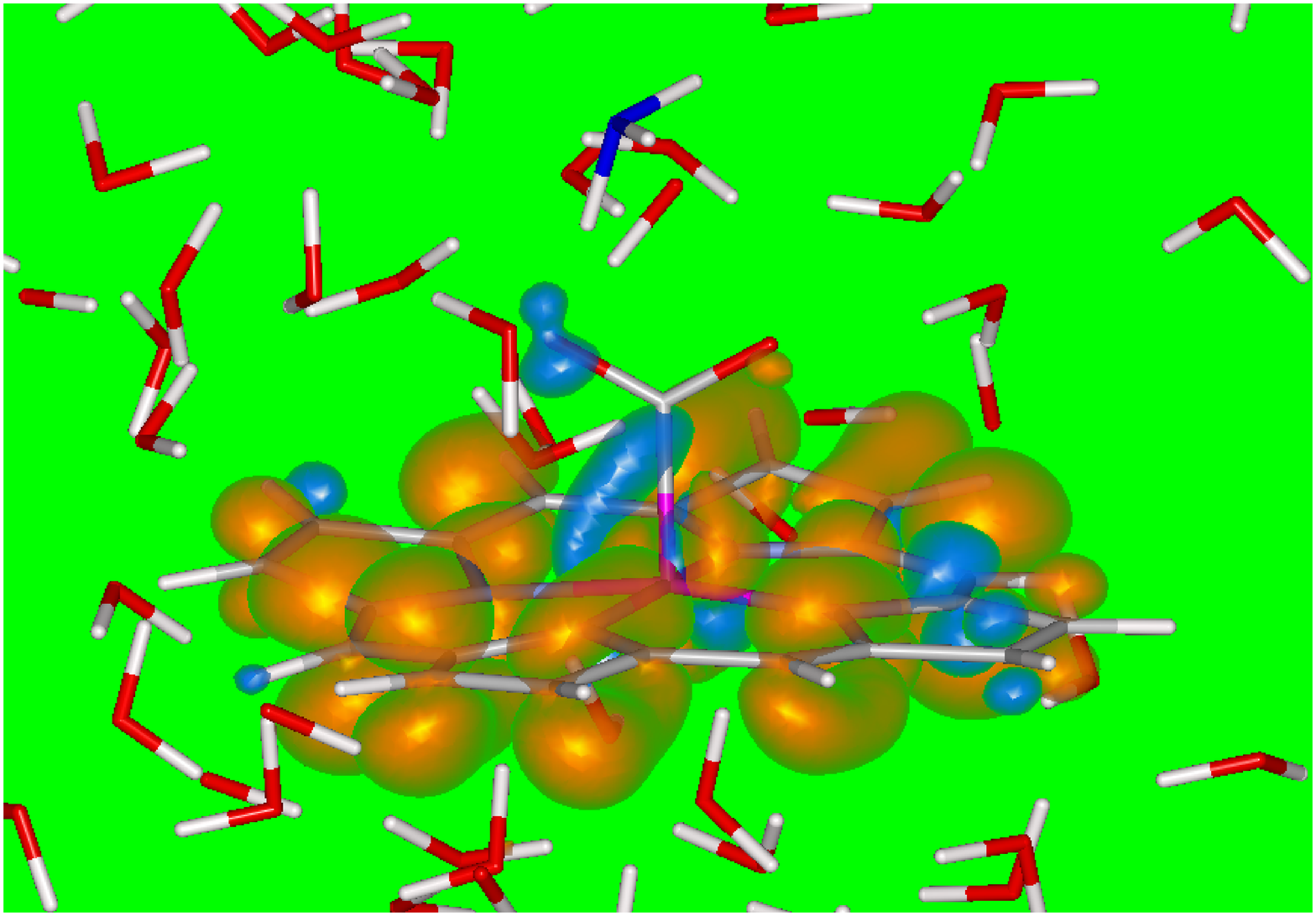}}
            \hbox{\epsfxsize=1.50in \epsfbox{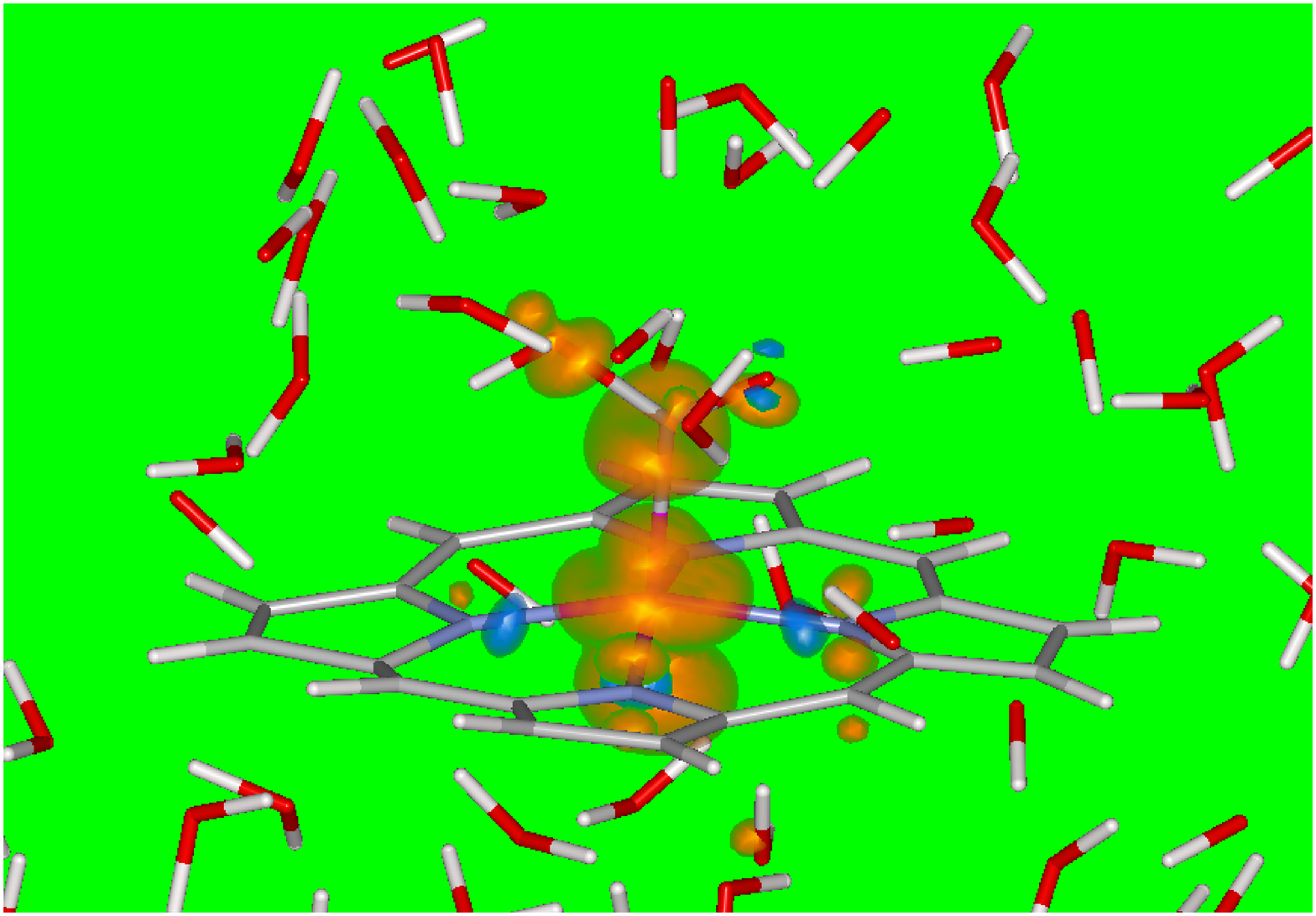}} (b)}
\centerline{(c) \hbox{\epsfxsize=1.50in \epsfbox{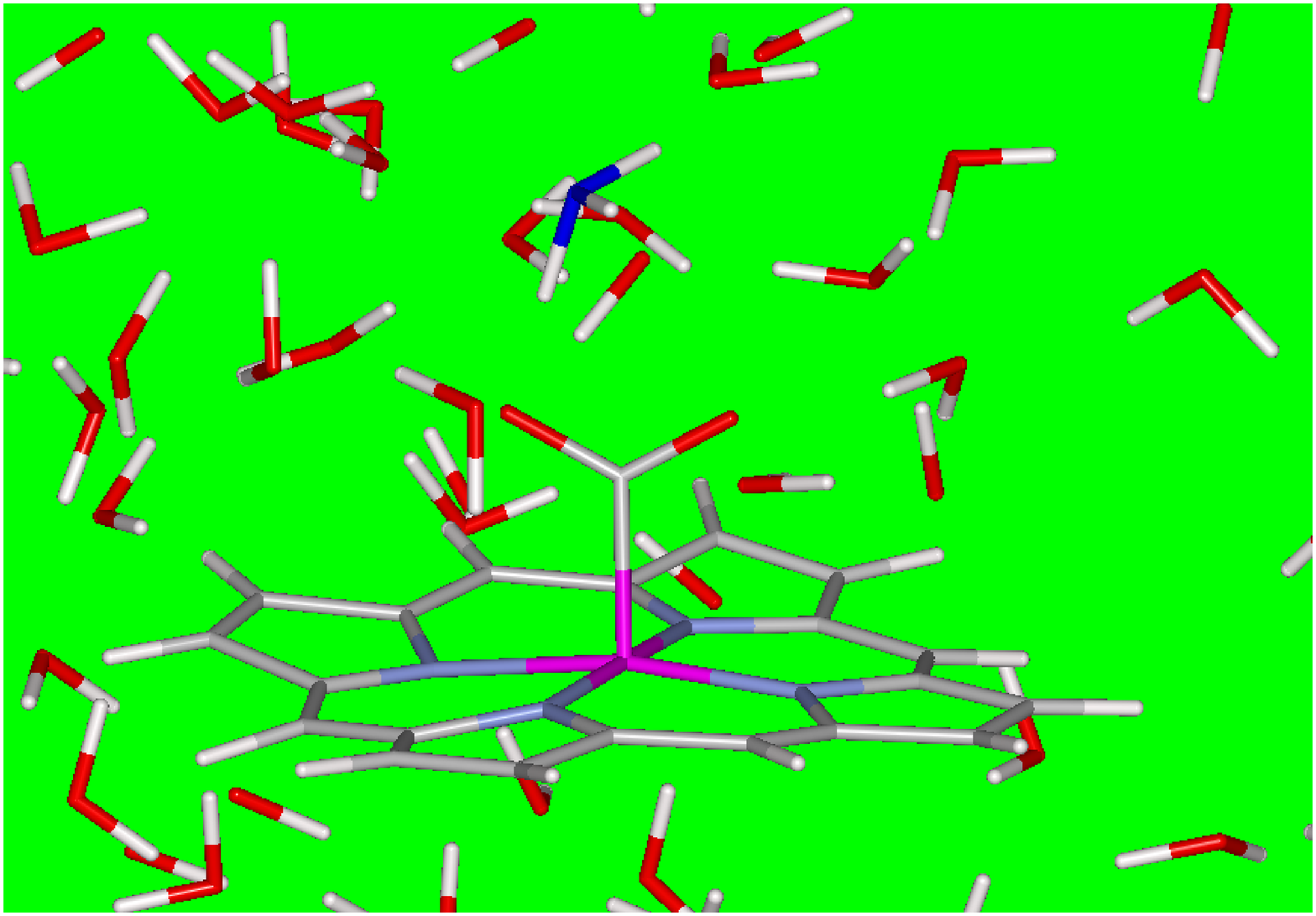}}
            \hbox{\epsfxsize=1.50in \epsfbox{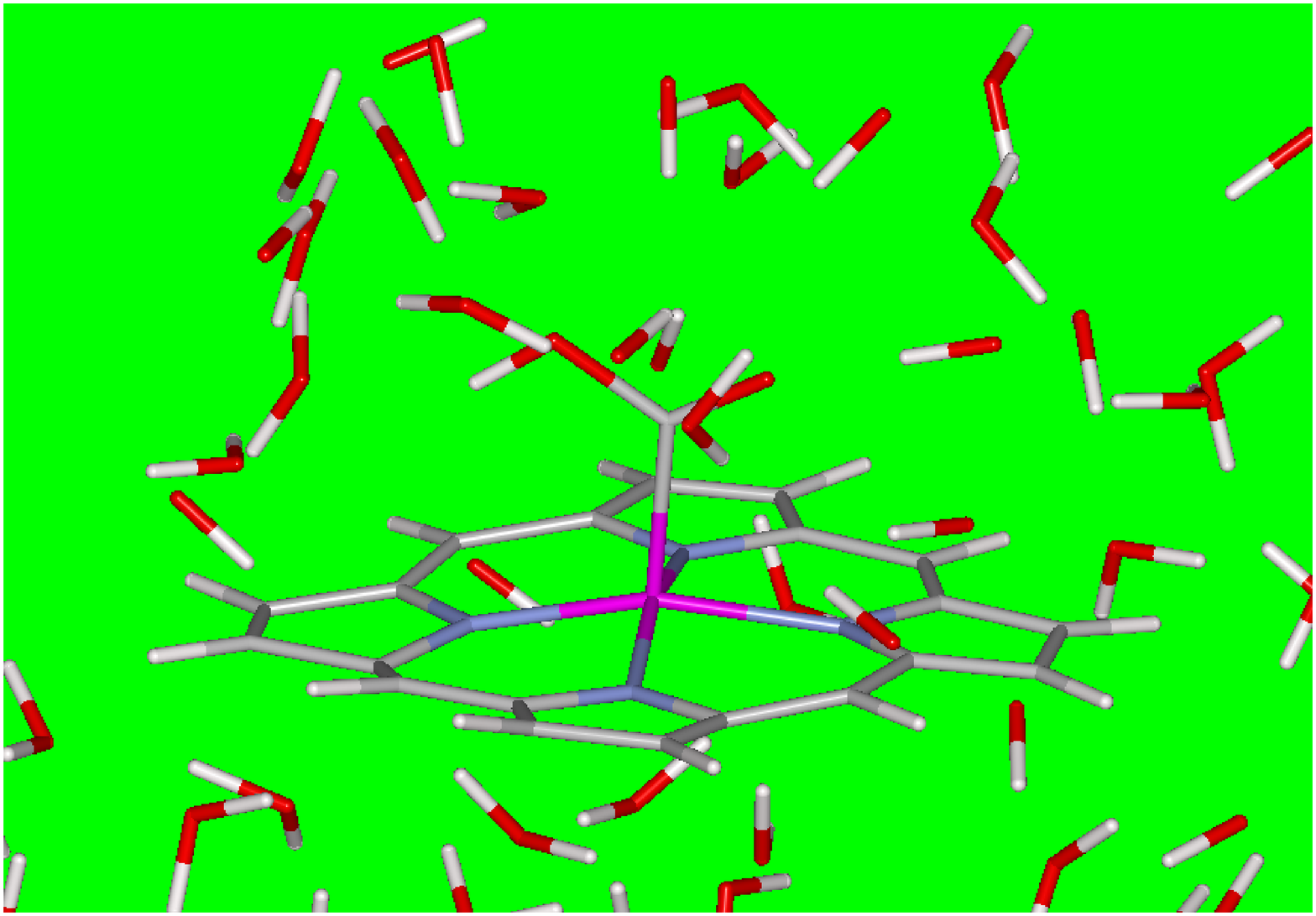}} (d)}
\centerline{(e) \hbox{\epsfxsize=1.50in \epsfbox{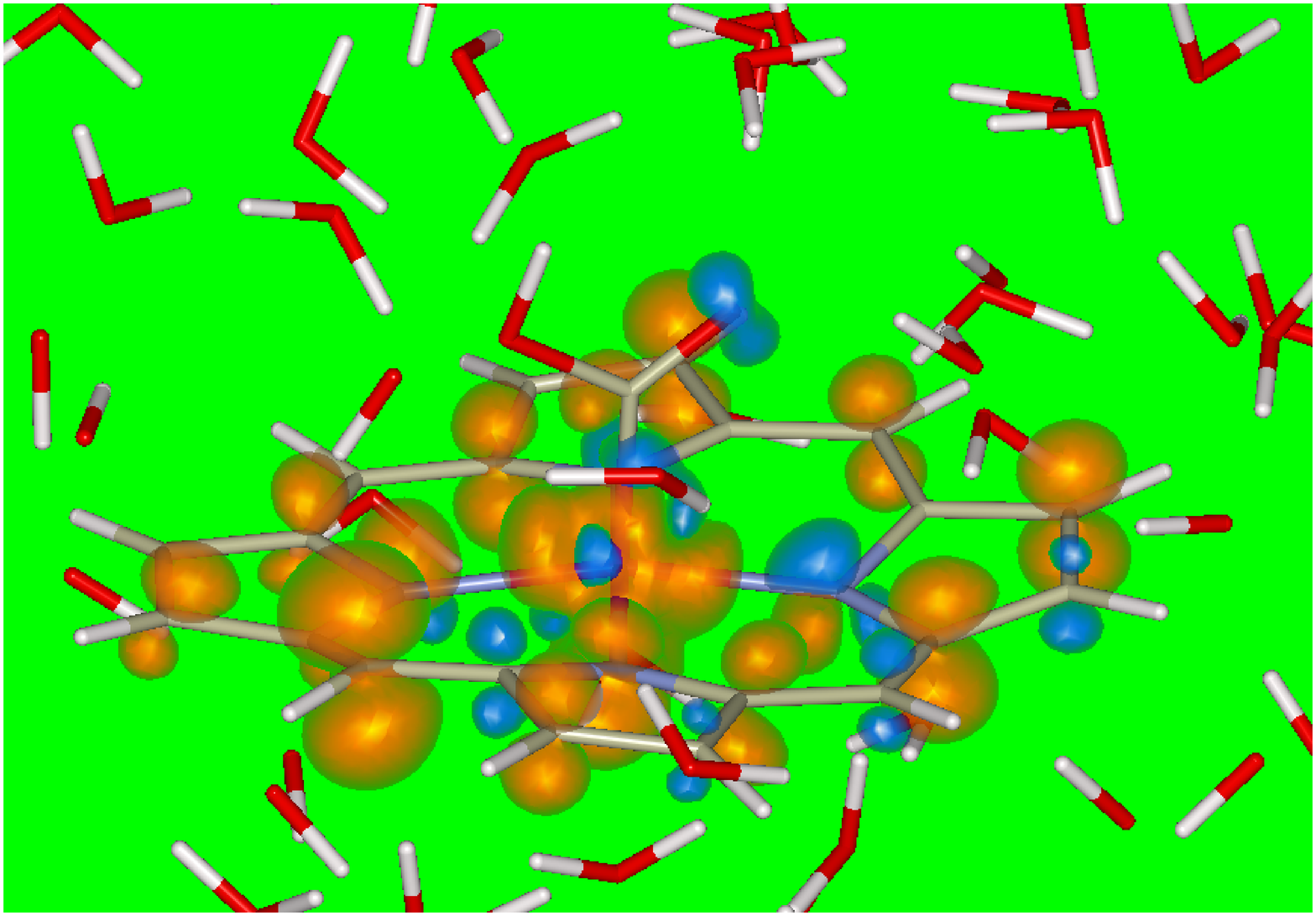}}
            \hbox{\epsfxsize=1.50in \epsfbox{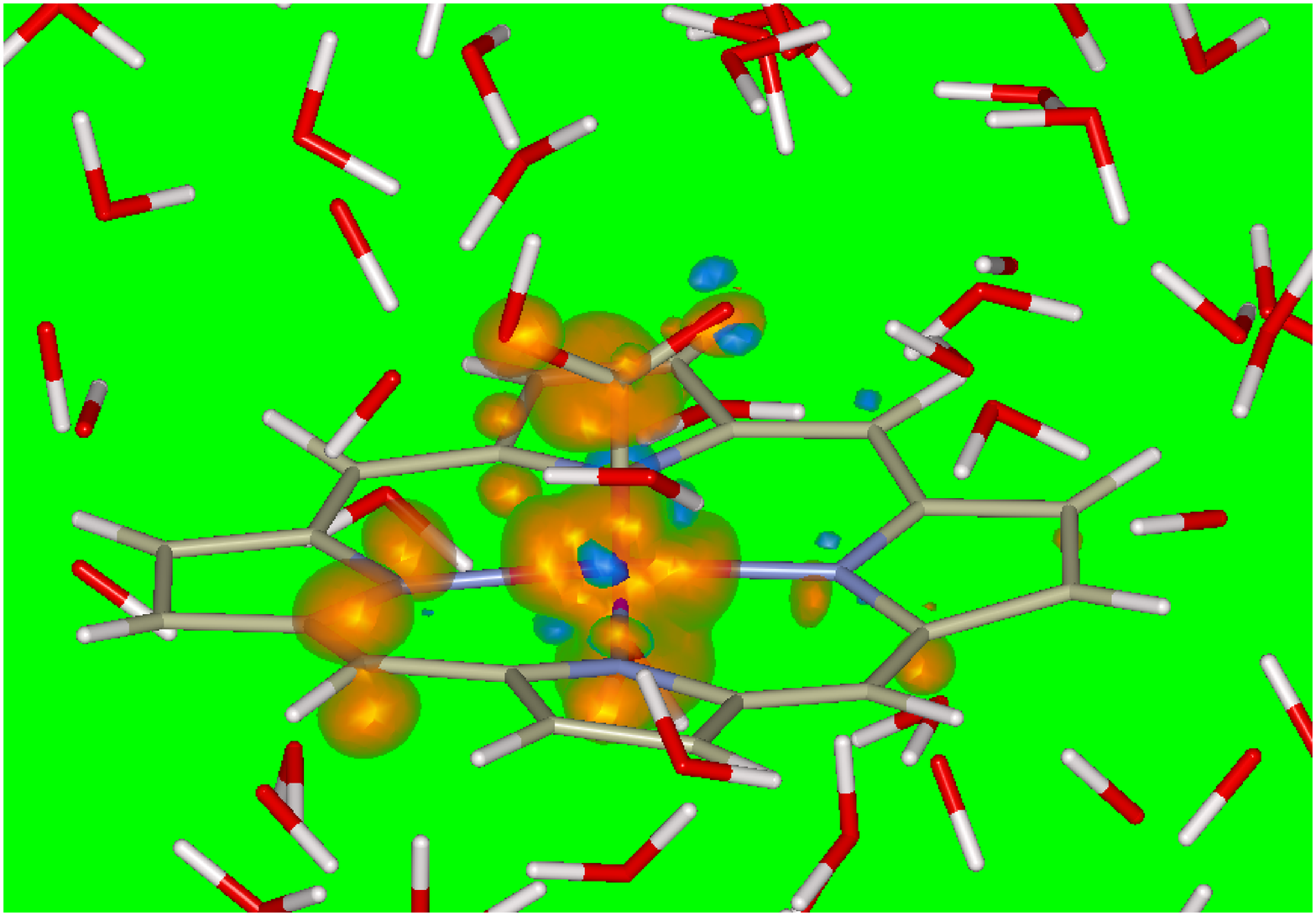}} (f)}
\caption[]
{\label{fig3} \noindent
Spin densities as protonation of [Co(I)PCO$_2$]$^{2-}$ proceeds.
Gold/blue: regions with net postive/negative spin densities.
The isosurface spin density values are different for all panels
to facilitate visualization, and are 1.32$\times$10$^{-4}$,
2.88$\times$10$^{-4}$, 2.32$\times$10$^{-4}$, and
2.72$\times$10$^{-4}$~$|e|$/\AA$^3$ for panels (a), (b),
(e), and (f), respectively.  The color scheme of the stick figures for
water and CoP is as in Fig.~\ref{fig1}.
(a) Well-equilibrated [Co(I)PCO$_2$]$^{2-}$:H$_3$O$^+$ contact ion pair,
obtained using a harmonic umbrella potential (Eq.~\ref{uofr}) with
$B$=3.0~eV/\AA, $R_o=-1.12$~\AA.  The singly occupied molecular orbital
(HOMO) state is delocalized on the porphyrin ring.  (b) [Co(II)PCOOH]$^-$,
$B$=3.0~eV/\AA, $R_o$=$-0.4$~\AA; the harmonic potential forces a H$_2$O
molecule to accept a hydrogen bond from the COOH proton.  (c) \& (d)
reprise panels (a) and (b), respectively, omitting the spin densities
to reveal the atomic positions more clearly.  The dark blue stick figure
above CoP in panel (c) represents the transient H$_3$O$^+$.  (e) \& (f):
Starting the trajectory from panel (a), $R_o$ is suddenly switched from
$-1.12$~\AA\, to $-0.4$~\AA.  The panels are taken 1.188875~ps and
1.189125~ps into this trajectory and bracket the transition.
}
\end{figure}

The Co charge and spin states predicted for B3LYP DFT+pcm
optimal structures\cite{paper1} and in DFT+U/AIMD aqueous
phase snapshots generally agree with each other.  For example,
we consider an AIMD snapshot of [Co(I)PCO$_2$]$^{2-}$:H$_3$O$^+$
contact ion pair\cite{silica} in water (Fig.~\ref{fig3}a, obtained using
Eq.~\ref{uofr} with $B$=3~eV/\AA$^2$ and $R_o=-1.12$~\AA.).\cite{graphics}
A maximally localized Wannier function analysis reveals that
6 occupied $d$-spin-orbitals are centered within 0.1~\AA\,
of the Co atom; another 2 occupied spin-orbitals are 0.86 to 0.88~\AA\,
away from the Co, and 1.04 and 1.02~\AA\, away from the CO$_2$ carbon
atom, respectively.  (The Co-C distance is 1.90~\AA\, in this snapshot.)
This electronic configuration is consistent with a dative covalent bond
donated by [Co(I)P]$^{2-}$ to CO$_2$.  The singly occupied highest
occupied HOMO state is delocalized on the porphyrin
ring, as has been discussed in Paper~I and is confirmed in the
spin density plot of Fig.~\ref{fig3}a.  We designate this species
[Co(I)PCO$_2$]$^{2-}$.  

A key exception to the agreement between AIMD and B3LYP DFT+pcm calculations
is [Co(II)PCOOH]$^-$.  Figure~\ref{fig3}b depicts an AIMD snapshot of this
species, obtained with harmonic constraint $R_o$=$-0.4$~\AA\, which yields
a COOH (i.e., protonated CO$_2$) group and also forces a H$_2$O molecule
to accept a hydrogen bond from the COOH proton.\cite{silica}
A Wannier analysis of this snapshot (Fig.~\ref{fig3}b)
reveals 7 occupied $d$-spin-orbitals centered around 0.2~\AA\, of the Co
atom.  Another two occupied spin-orbitals are localized along the Co-C bond,
within 0.6 and 0.8~\AA\, of the C-atom respectively.   These Wannier orbital
centers are closer to C than Co, and the electronic configuration is
consistent with a [COOH]$^-$ group attached to a [Co(II)P]$^0$.
The HOMO state is a Co $d$-orbital; the spin density of this species
(Fig.~\ref{fig3}b) is far more localized than that of
[Co(I)PCO$_2$]$^{2-}$ (Fig~\ref{fig3}a).  In contrast, B3LYP calculations
with dielectric approximation for water suggest a [Co(I)P]$^{2-}$-COOH$^+$
electronic configuration, and the HOMO is a $\pi$-orbital on the porphyrin
ring there.\cite{paper1}  This difference most likely reflects the explicit
treatment of molecular water in the DFT+U/AIMD simulation which helps to
stabilize a COOH$^-$ group via hydrogen bonding.  In the SI, Sec.~S5,
explicit treatment of water molecules in the first hydration shell
of the COOH group is indeed shown to yield a Co(II) charge state.

The change of electronic structure coupled to protonation of 
[Co(I)PCO$_2$]$^{2-}$ has consequences for AIMD simulations.
Figures~\ref{fig3}e-f depict the spin density transition in real time,
occurring about 1.2~ps after the harmonic umbrella sampling potential
(Eq.~\ref{uofr}) is suddenly switched from $R_o=-1.12$~\AA\ (consistent with
that in Fig.~\ref{fig3}a) to $R_o$=$-0.4$~\AA\,
(Fig.~\ref{fig3}b).  As alluded to above, this change in
$R_o$ leads to reprotonation of [Co(I)PCO$_2$]$^{-2}$ from its H$_3$O$^+$ 
neighbor and the motion of a water molecule towards the CO$_2^-$ group.
Simultaneously, the electronic structure relaxes to [Co(II)PCOOH]$^-$
due to the large driving force arising from the nuclear motion.
However, if the driving force is not sufficient, e.g., if $R_o$ is switched
to an intermediate $-0.7$~\AA, the system may take much longer than 1.2~ps
to spontaneously sample the Co(II) charge state.  As discussed below,
this leads to a slight hysteresis in the umbrella sampling calculation.

\subsection{Protonation of [CoPCO$_2$]$^{2-}$}

\begin{figure}
\centerline{\hbox{\epsfxsize=4.5in \epsfbox{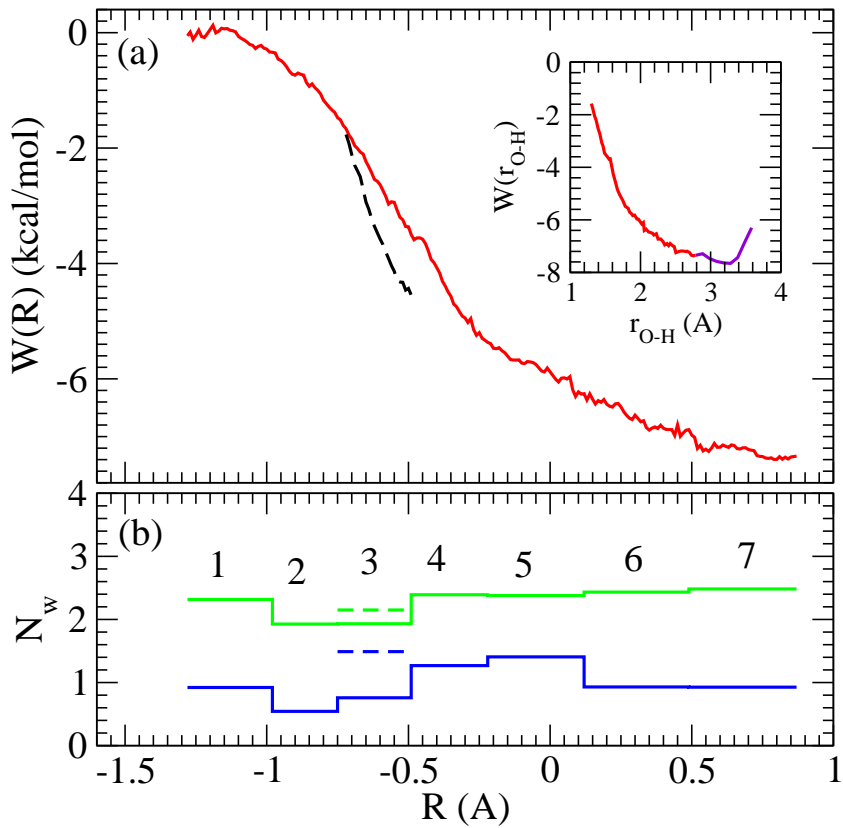}}}
\caption[]
{\label{fig4} \noindent
(a) Red: Potential of mean force ($W(R)$) associated with [Co(II)PCOOH]$^-$
deprotonation.  The most negative values of $R$ are associated with
complete deprotonation while $R>-0.4$~\AA\, refers to the protonated state.
Inset: the violet line depicts the ${\bar W}(r_{\rm O-H})$, computed using
unbiased AIMD simulations; the red line depicts ${\bar W}(r_{\rm O-H})$
transformed from the biased (umbrella sampling) $W(R)$ and is a cubic spline
fit.  (b) Hydration numbers ($N_w$) of the carbonyl (green) and hydroxyl
(blue) oxygen atoms on the COOH group. $N_w$ is defined as the number of
water protons within 2.5~\AA\, of these oxygen sites.
Cobalt is in the Co(I) state in windows 1-3 and Co(II) state in 4-7,
except for the dashed curves (window~3) where Co(II) prevails.
The dashed lines shows that window~3 is affected by 
the cobalt charge state hysteresis (see text).
}
\end{figure}

Figure~\ref{fig4}a depicts the $W(R)$ associated with
the deprotonation reaction
\begin{equation}
[{\rm Co(I)PCO_2}]^{2-} + {\rm H}^+ \rightarrow [{\rm Co(II)PCOOH}]^- .
		 \label{eq5} 
\end{equation}
Our 4-atom coordinate $R$ effectively
interpolates between the large negative $R$ deprotonated plateau region
where $W(R)\rightarrow 0$, related to water-separated ion pairs,\cite{silica}
and the protonated $R > -0.4$~\AA\, region where the shallow curvature
of $W(R)$ is governed by hydrogen bonding between the acid proton and
a water molecule serving as a hydrogen bond acceptor.  The distribution of
shortest distance between a water oxygen and the COOH acid proton, obtained
in an unconstrained (i.e., $B=0$ in the Eq.~\ref{uofr} $U(r)$)
AIMD simulation of [Co(II)PCOOH]$^-$ in water, is
expressed as a free energy profile (${\bar W}(r_{\rm O-H})$)
in the inset of Fig.~\ref{fig4}a.  The optimal $r_{\rm O-H}$
is 3.2~\AA.  Recalling the definition Eq.~\ref{R_def} and the fact
that $R_2$ and $R_3$ are O-H covalent bonds of lengths $\sim 1$~\AA, this
optimal value implies that a true minimum in $W(R)$ should not emerge until
$R \sim 1.2$~\AA.  The optimal $r_{\rm O-H}$ distance is larger than
the canonical hydrogen bond cutoff distance
of 2.5~\AA, and reflects the inability of the COOH proton, engaged in a
strong intramolecular hydrogen bond to one of the CoP nitrogen atoms, to
donate a hydrogen bond to water molecules (Fig.~\ref{fig1}f).  Indeed,
along the AIMD trajectory, there is only a 2~\% probability that the
COOH proton and any water oxygen atoms are within 2.5~\AA\, of each other.
Fortunately, the reaction coordinate $R$ and the umbrella constraining
potentials enforce hydrogen bond donation from the acid proton to water,
a pre-requisite to deprotonation.

To estimate p$K_{\rm a}$ via Eq.~\ref{eq5}, we find the most probable
O$_{\rm water}$-H$^+$ hydrogen bond distance $r_{\rm O-H}$
at each $R$, thus locally converting $W(R)$ to $\bar{W}(r_{\rm O-H})$,
and perform a spline fit to that probability
distribution.\cite{sprik,silica}  The result is 
matched to ${\bar W}(r_{\rm O-H})$ in the small $r_{\rm O-H}$ region
obtained in the aforementioned AIMD run where the umbrella potential is
absent.  See the Fig.~\ref{fig4}a inset.  Integrating over $r_{\rm O-H}$
with a $4\pi r^2_{\rm O-H}$ volume element, which takes the place of the
phase space factor $A(R)$ in Eq.~\ref{eq8}, referencing to water
auto-ioniziation computed at a similar elevated temperature,\cite{silica}
and adding a $-0.57$~kcal/mol zero point energy correction estimated
from gas phase B3LYP calculations, p$K_{\rm a}$=9.0$\pm$0.4 is predicted.
Thus, [Co(II)PCOOH]$^-$ does not behave like an ordinary carboxylate acid with
p$K_{\rm a} \sim 4.5$.  The significant reduction of acidity
indicates that protonation of [Co(I)CO$_2$]$^{2-}$ is exothermic
at the experimental pH~$> 7$ conditions.

In contrast, a preliminary study of the deprotonation $W(R)$ of
Co(III)PCOOH reveals that $W(R)$ is almost independent of $R$ for
$R < -0.4$~\AA\, (not shown).  Our reaction coordinate has been used
to calculate p$K_{\rm a}$ down to 3.8 where a finite curvature
persists in $W(R)$.\cite{silica}  This suggests that the 
p$K_{\rm a}$ of Co(II)PCOOH is less than 3.8, much lower than
that of [Co(II)PCOOH]$^-$.  Thus, adding an electron to the CO$_2$-ligated
catalyst evidently enhances its ability to hold on to excess
protons.\cite{brookhaven}  While this may appear obvious in retrospect,
the strong intramolecular hydrogen bonding in [Co(II)COOH]$^-$
(Fig.~\ref{fig1}f) also likely contributes to its higher p$K_{\rm a}$.  This
preliminary p$K_a$ estimate for Co(III)PCOOH suggests that protonation of
CoPCOO$^-$ cannot occur spontaneously at the experimental pH$\sim$7, further
confirming the B3LYP DFT+pcm $\Phi_{\rm redox}$ prediction that
[Co(II)PCOOH]$^{-}$,
not [Co(III)PCOOH], is the key intermediate (Fig.~\ref{fig2}).

Figure~\ref{fig4}b shows that the hydration numbers $N_w$ do not
significantly vary with $R$.  However, the hydration structure may
determine whether hysteresis in
the cobalt charge state occurs as $R$ increases.  The full curve in
Fig.~\ref{fig4} collates contributions from all windows, such that
windows~1 and~3 are initiated from [Co(I)PCO$_2$]$^{2-}$ configurations
generated in a well-equilibrated window~2 trajectory, while the window~5-7
runs originate from window~4 ([Co(II)PCOOH]$^-$).  This appears to
yield a smooth $W(R)$ curve.  If the window 3 segment of $W(R)$ is
initiated from window~4 instead, the system remains in a Co(II) state,
and the dashed curve, matching poorly to the window~2, materializes.
This is a classic signature of hysteresis in umbrella sampling
simulations.  Fortunately, only the intermediate $R_o=-0.7$~\AA\,
window~3 suffers from this problem.  The maximal underestimation of
[Co(II)PCOOH]$^-$ p$K_a$ this can introduce is the difference between the
full and dashed curves in window~3, which is 0.76~kcal/mol or 0.55~pH unit.  
This hysteresis issue is discussed in more detail in the appendix.

Using a purely dielectric continuum treatment of water and the
B3LYP/6-31+G$^*$ method, Paper~I predicts deprotonation p$K_{\rm a}$
of 13.8 and 7.6 for [Co(II)PCOOH]$^-$ and Co(III)PCOOH, respectively.
They are several pH units higher than AIMD estimates.  The discrepancies
are partly due to the lack of explicit water molecules in the B3LYP
DFT+pcm calculations, but they also reflect the substantial ($\sim 5$~kcal/mol)
variation in deprotonation free energies when using different
DFT functionals (Table~4 in Paper~I).  

\subsection{C-OH bond Cleavage Reaction}

\begin{figure}
\centerline{\hbox{\epsfxsize=4.5in \epsfbox{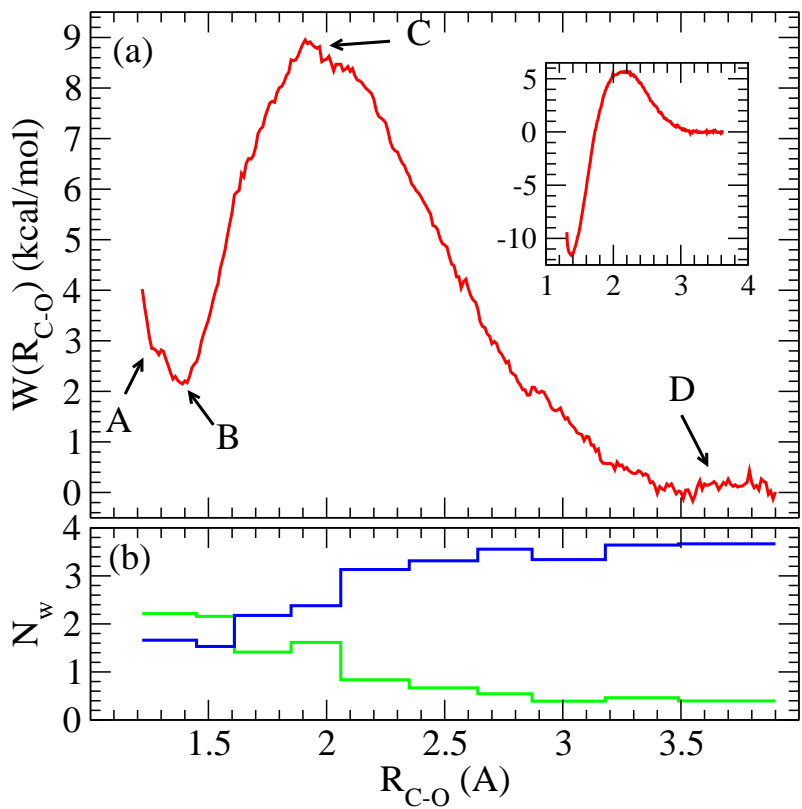}}}
\caption[]
{\label{fig5} \noindent
(a) $W(R)$ for the C-OH cleavage reaction, Eq.~\ref{eq6}.   Inset: $W(R)$
for CO$_3$H$^-$ $\rightarrow$ CO$_2$ + OH$^-$ from Ref.~\onlinecite{co2}.
Snapshots at points A-D are depicted in Fig.~\ref{fig6}. (b) Hydration
numbers ($N_w$) for the carbonyl (green) and hydroxyl (blue)
oxygen atoms as $R_{\rm C-O}$ varies.}
\end{figure}

Finally, we apply umbrella sampling to study
\begin{equation}
{\rm [Co(II)PCOOH]}^- \rightarrow {\rm [Co(II)PCO]} + {\rm OH}^- .\label{eq6}
\end{equation}
We use the C-O distance of the C-OH bond as the reaction coordinate. 
Figure~\ref{fig5}a shows that this reaction is exothermic.  Wannier analysis
reveals that the system remains in the Co(II) state as $R_{\rm C-O}$ varies,
and no hysteresis is observed.  After accounting for standard
state correction, the $4\pi R_{\rm C-O}^2$ rotational contribution,
a $-2.5$~kcal/mol ZPE correction, and integrating
$\exp [-\beta W(R_{\rm C-O})]$ in the reactant channel as in Eq.~\ref{eq8},
we obtain a free energy of reaction $\Delta G^{(0)}=-8.5\pm$1.1~kcal/mol.
The barrier height is a low $\Delta G^{(0)*}=5.2\pm$0.6~kcal/mol confirms
that the activation free energy is fairly low.  We have not attempted
to compute ZPE for $\Delta G^{(0)*}$ which requires a projecton 
operation that removes the reaction coordinate;\cite{merz} however,
in a study of a simple C-OH bond breaking reaction,\cite{merz}
ZPE has been found to be small, reducing $\Delta G^{(0)*}$ by 0.8~kcal/mol.

The predicted activation barrier may depend on the
$R_{\rm CO}$ coordinate chosen.  As in Ref.~\onlinecite{co2},
we have computed the transmission coefficient.\cite{book1}
Thus, in the umbrella sampling window containing the $W(R)$ turning
point, we randomly choose 10 configurations at the top of the barrier,
half with positive velocities $d R_{\rm C-O} /dt$ and half with negative
ones, restart AIMD trajectories without umbrella sampling potentials,
and determine the ratio $\kappa$ that the reaction proceed without ultimate
recrossing back to the reactants. 
$\kappa=1$ means no recrossing and a perfectly chosen reaction coordinate.
We find that $\kappa=0.60$, indicating that $R_{\rm CO}$ is a reasonable
coordinate and that our reported $\Delta G^*$ should be qualitatively correct.
A more systematic approach is the path-sampling method,\cite{chandler}
which is however more computationally costly.  

The C-OH cleavage activation barrier is almost a factor of 3 smaller than
that previously found for the uncatalyzed CO$_3$H$^-$ $\rightarrow$ CO$_2$
+ OH$^-$ reaction (Fig.~\ref{fig5}a inset).\cite{co2}  Even though the
comparison is not perfect, in that the carbon atom is not reduced to its
+2 oxidation state in the previous work, the cobalt porphyrin
has clearly and drastically reduced the C-OH cleavage barrier.  

\begin{figure}
\centerline{(a) \hbox{\epsfxsize=1.50in \epsfbox{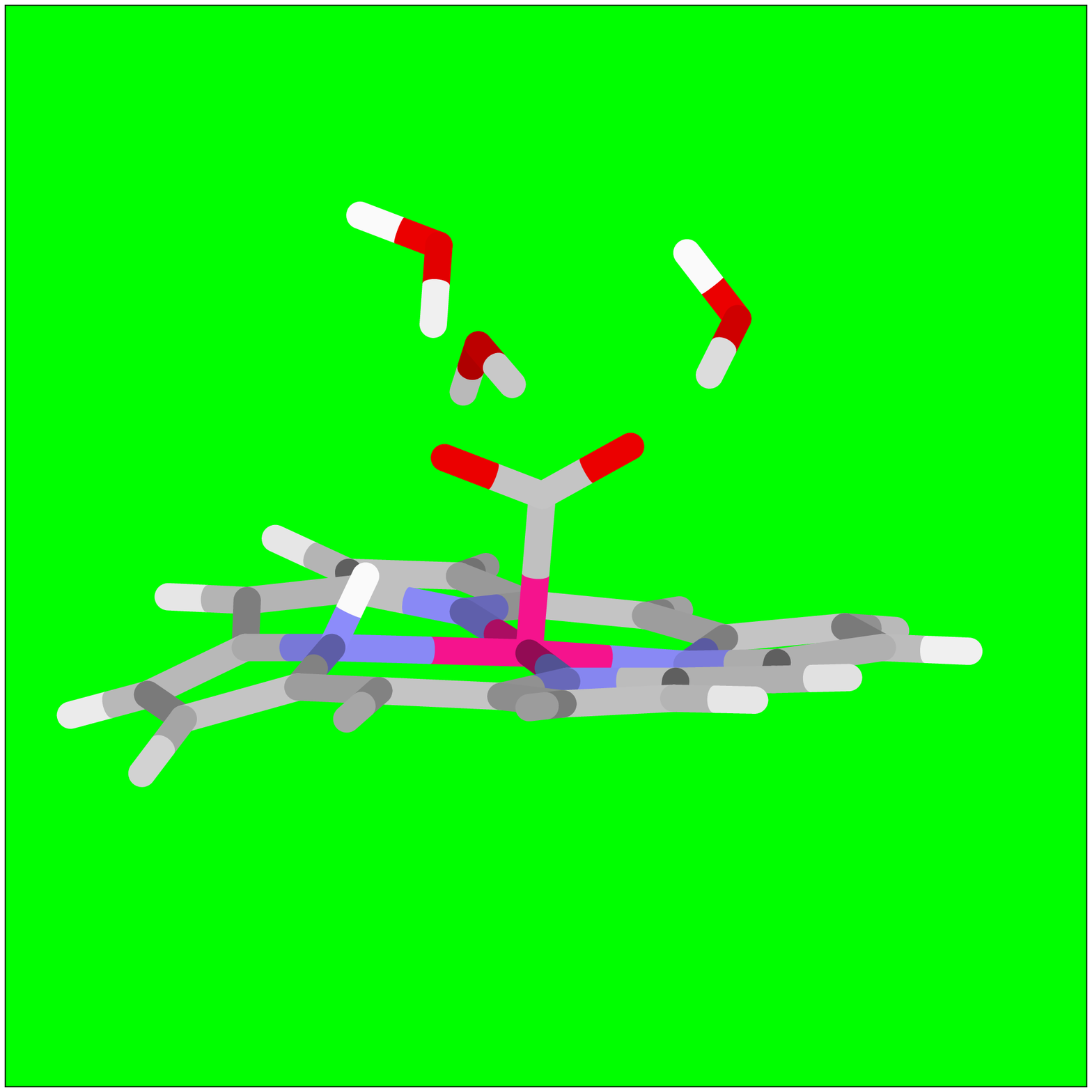}}
            \hbox{\epsfxsize=1.50in \epsfbox{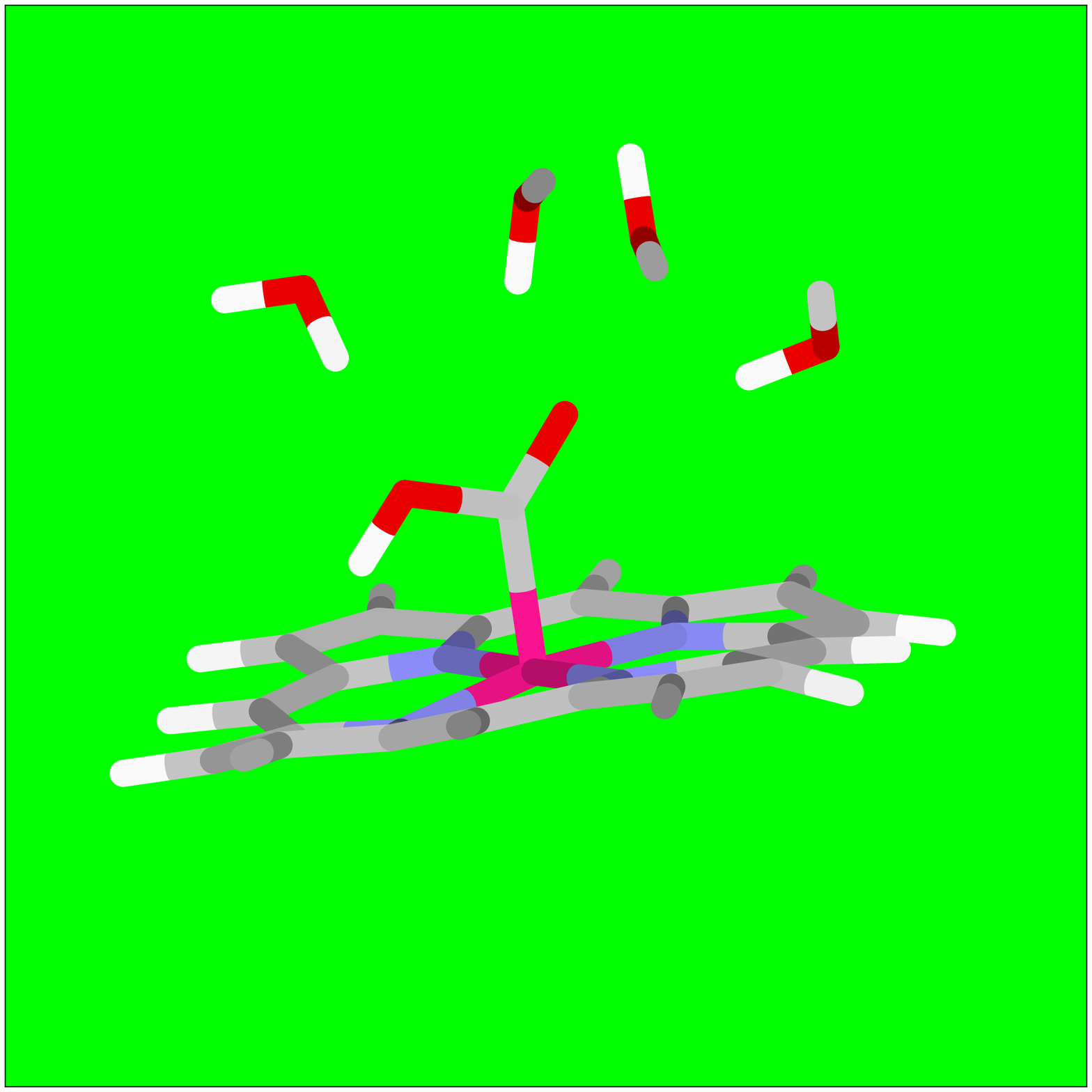}} (b)}
\centerline{(c) \hbox{\epsfxsize=1.50in \epsfbox{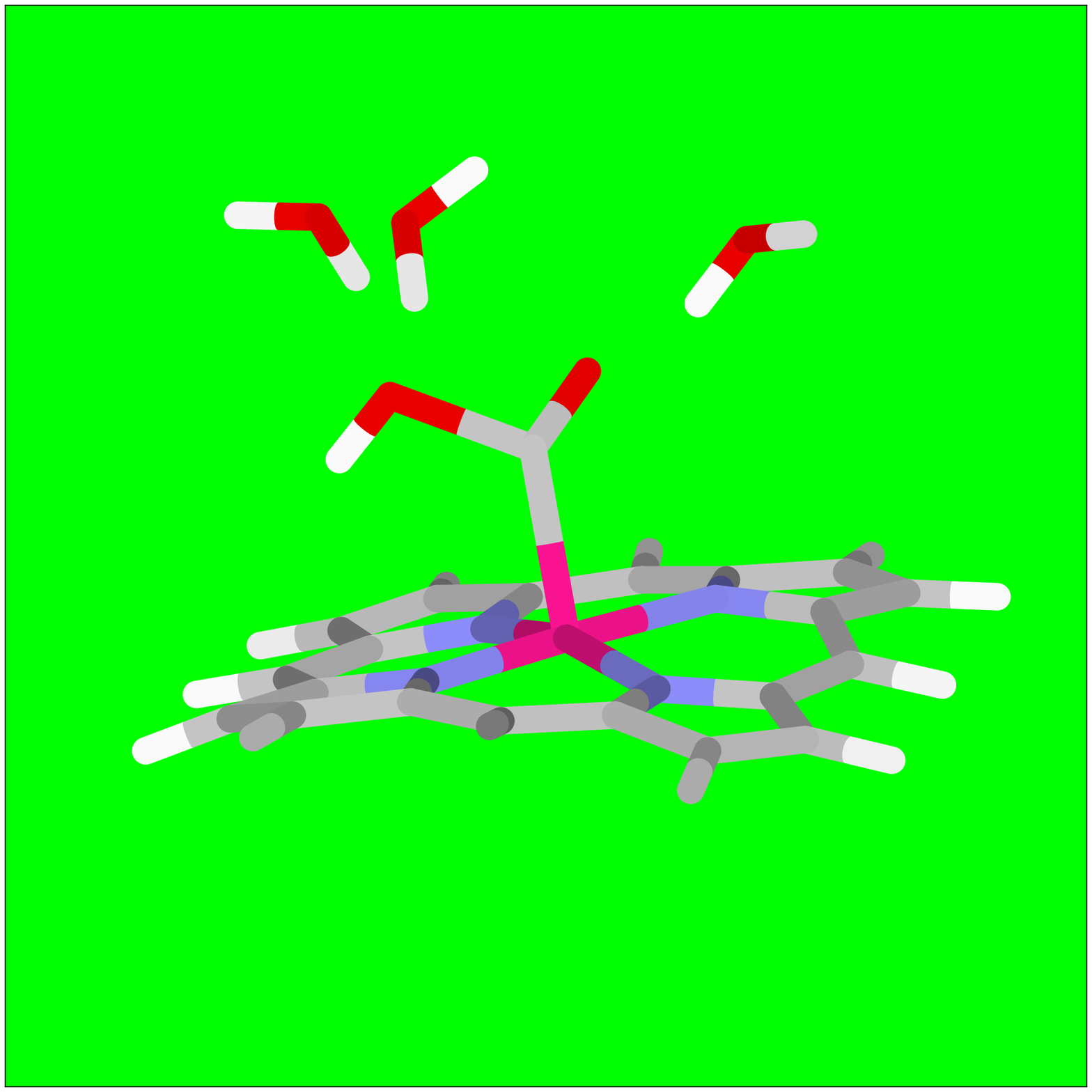}}
            \hbox{\epsfxsize=1.50in \epsfbox{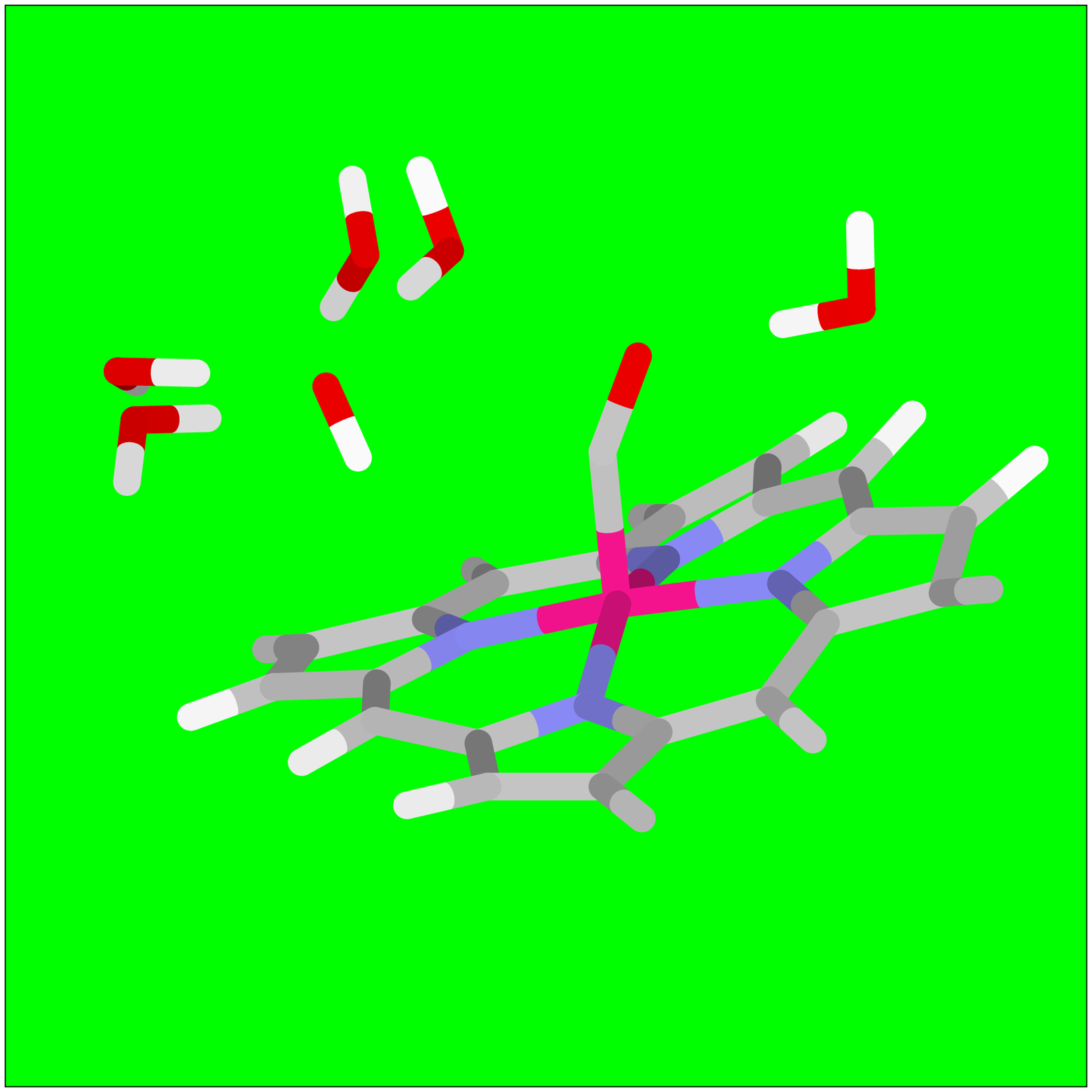}} (d)}
\caption[]
{\label{fig6} \noindent
(a)-(d): snapshots in the four umbrella sampling windows indicated in
Fig.~\ref{fig5}.
AIMD simulations are conducted in an explicit liquid water environment;
only a few water molecules are shown in these snapshots.  In panel (a),
the proton on the COOH group has migrated to one of the nitrogen
atoms on the porphyrin ring.  The color scheme is as in Fig.~\ref{fig1}.
}
\end{figure}

As the C-OH cleavage reaction proceeds, the carbonyl oxygen in the
initially partially negatively charged COOH$^-$ functional group
becomes part of an uncharged carbon monoxide molecule (CO) weakly bound
to Co(II)P, and the oxygen atom of the nascent CO exhibits a hydration
number $N_w$ which steadily decreases (Fig.~\ref{fig5}b).
In contrast, the hydroxyl oxygen transitions towards a hydroxide
anion (OH$^-$), and its $N_w$ increases to about 3.5.  The $N_w$ value for 
the emerging OH$^-$ oxygen in the present heterogeneous environment is 
therefore similar to that predicted in bulk liquid water using the PBE
functional.\cite{oh}  Figure~\ref{fig6} further depicts snapshots of the
instantaneous hydration structure of the COOH group at different
values of the reaction coordinate.  Panel (b) represents a configuration
where the COOH proton is intramolecularly hydrogen bonded to a porphine
ring N atom (Fig.~\ref{fig1}f).  In panel (a), which corresponds to a
kink in Fig.~\ref{fig5}a, the COOH proton has instantaneously been
donated to a N atom, forming a covalent bond with it.  In the gas phase,
this N-H bonded structure is 6.68~kcal/mol higher in energy
than that of Fig.~\ref{fig6}b.  Nevertheless, in the aqueous phase, 
this configuration is occasionally observed.  

\section{Discussion}
\label{discussion}

\subsection{Comparison with Previous CO$_2$ Theoretical Work}

In this subsection, we make comparisons with some computational
aspects of Ref.~\onlinecite{co2} and with Paper~I.\cite{paper1} 


Unlike Ref.~\onlinecite{co2}, we have not constrained the OH bond rotation
around the C-O axis in the cleaved COOH group and then corrected for the
entropic contributions there.  This is because the PBE functional we use is
consistent with much faster OH$^-$ dynamics in water than the RPBE functional
previously applied,\cite{co2,rpbe,oh} and it is reasonable to assume that
the OH rotation around the C-O axis is better-sampled than in RPBE simulations
within 10~ps AIMD trajectories.  This should only affect $\Delta G^{(0)}$,
not $\Delta G^{(0)*}$, because the C-OH bond is not completely broken at
the transition state at $R_{\rm C-O}=1.9$~\AA\, and free OH rotation around
the C-O axis does not occur there.  The important qualitative conclusion of
this work is that $\Delta G^{(0)}$ is exothermic and that C-OH bond breaking
is thermodynamically downhill; the precise free energy
change associated with Eq.~\ref{eq6} is less important.

We have not attempted to correct the AIMD/DFT+U $W(R)$ with single
point MP2 calculations as was done in Ref.~\onlinecite{co2}.  The
CoP systems examined in this work are too large for large-basis MP2.
Furthermore, in Ref.~\onlinecite{co2}, the DFT functional used was
RPBE\cite{rpbe} which is arguably less accurate for heterogeneous
C-O bond breaking than the PBE functional used herein.

Unlike Paper~I, we have focused on Eq.~\ref{eq3} and not the
proton-assisted variation
\begin{eqnarray}
{\rm [Co(II)PCOOH]}^- + {\rm H}^+ &\rightarrow& 
	{\rm Co(II)PCO}({\rm OH}_2) ; \nonumber \\
{\rm Co(II)PCO}({\rm OH}_2) &\rightarrow& {\rm [Co(II)PCO]}
			+ {\rm H}_2{\rm O} .\label{eq10}
\end{eqnarray}
Thermodynamically the two are equivalent.\cite{note98}  In terms of
kinetics, which Paper I does not deal with, they will have different
activation barriers.  To estimate
$\Delta G^{(0)*}$ for Eq.~\ref{eq10}, we note that all AIMD $W(R)$
reported for protonation/deprotonation reactions in the literature
have been monotonic, i.e., the free energy changes and activation
barriers are the same.  Assuming the second half of Eq.~\ref{eq10} is
also fast to the point of being barrierless, which is a lower limit on
the overall Eq.~\ref{eq10} barrier, $\Delta G^{(0)*}$ for Eq.~\ref{eq4}
would be entirely due to
the $\Delta G^{(0)}$ of the first half of this equation and related to the
p$K_{\rm a}$ of Co(II)PCO(OH$_2$).  At the pH=7 experimental conditions,
the p$K_{\rm a}$ of the C-OH$_2$ group in Co(II)PCO-OH$_2$ will have
to be above 3.2 in order to have a lower barrier than the
$\Delta G^*=5.2$~kcal/mol we find for Eq.~\ref{eq3}.  This appears
unlikely; C-OH$_2$ groups tend to be very acidic and lose protons readily.
In any case, if this alternate, proton assisted route of C-O cleavage
were faster, the barrier estimated in our PMF calculation (Fig.~\ref{fig5})
would be an upper bound to the reaction activation free energy; the C-O
bond breaking step would still exhibit fast dynamics, and the qualitative
conclusion of this paper would be unchanged.  We plan to
revisit Eq.~\ref{eq10} in the future.

\subsection{Alternative Computational Methods}

A two-dimensional PMF calculation should remove the hysteresis
behavior in the protonation reaction (Eq.~\ref{eq6}, Fig.~\ref{fig4}).
A convenient second variable
may be the hydration number of the O~atom in the OH group.  This is
because the black dashed curve in Fig.~\ref{fig4}a exhibits $N_w$=1.49 for
that oxygen, considerably higher than the $N_w$=0.75 for the 
red, well-behaved $W(R)$ segment, suggesting that the electronic structure
is correlated with the average hydration number.  2-D PMF simulations would be
substantially accelerated using the metadynamics method\cite{meta1,meta2,meta3}
over traditional umbrella sampling.

Our p$K_{\rm a}$ calculations suggest that deprotonation of weak acid groups
which exhibit intramolecular hydrogen bonding and do not donate hydrogen
bond to water may require the use of a large number of
sampling windows.  This difficulty may be circumvented
by reversibly annihilating the proton using
an artificial reaction pathway\cite{sprik_new_proton} in a what might be
called a ``molecular grand canonical Monte Carlo'' approach.\cite{anatole}
This method does introduce the disadvantage of changing the net charge in
the finite-sized simulation cell, and may require a number of new
conformational constraints.

Finally, the self-consistent DFT+U approach\cite{cocc} has been
tested for Eq.~\ref{eq7}.   This method has the potential to establish
a $U$ value without resorting to parameterization with experimental
results.  Our preliminary studies suggest that this approach yields
a Co(III)P-CO binding energy that is too small, but
further development of this promising approach is under way.\cite{cocc2}

The above theoretical considerations, touching on many newly developed
techniques, emphasize the complexity and challenges associated with
modeling electrochemical reactions in explicit-water aqueous phase simulations.

\subsection{Accuracy of DFT functionals and redox potentials}

Accurate DFT functionals and dielectric continuum approximation of
the aqueous solvent are critical for DFT+pcm determination of redox
potentials,\cite{b3lyp_redox,truhlar04,redox2,truhlar} which in turn
govern the viable reaction intermediates and the overall reaction mechanism.
We have so far considered PBE, B3LYP, and DFT+U electronic structure
methods with the $U$ parameter in DFT+U fitted to an experimental
binding constant (Table~\ref{table1}).  B3LYP and DFT+U redox potentials
track each other and should predict the same reaction mechanism
at voltages slightly more negative than the Co(I)P/Co(II)P couple.
In contrast, the PBE functional predicts that [Co(II)COOH]$^-$ is extremely
unstable with respect to the far more acidic [Co(III)COOH] complex near the
PBE Co(I)P/Co(II)P redox potential.  The B3LYP $\Phi_{\rm redox}$ predictions
are more consistent with AIMD hydration structure considerations and
the fact that the CO$_2$-reduction reaction readily occurs near neutral
pH in a carbonate buffer (Sec.~\ref{results}).

More accurate DFT functionals may be used in the future to further
examine the mechanism proposed herein.  Candidates include the M06
class of functionals designed to yield better thermochemistry
accuracy for transition metal complexes,\cite{truhlar_tm1}
the B4(XQ3)LYP functional which has been found to yield improved
redox potentials for a suite of test cases,\cite{redox2} and
Gutzwiller wavefunction based methods.\cite{gutz}
The quality of the PCM dielectric continuum model\cite{pcm}
should also be further examined.\cite{b3lyp_redox}

\section{Conclusions}

In this work, we have applied first principles calculations to examine
the mechanism of the multi-step, two-electron electrochemical reduction
of CO$_2$ to CO in water using cobalt porphyrin (CoP) as catalyst.
First we have extracted redox potentials from DFT plus dielectric
continuum solvation calculations
using the B3LYP functional and a dielectric continuum treatment of
water.\cite{paper1} Even though the absolute value for the
[Co(I)P]$^-$/Co(II)P couple is not in good agreement with experiments,
the relative values of various redox potentials allow us to determine
where the electron transfers occur among the four intermediate steps.
Due to the enhanced interaction of CO$_2$ with water when bound to
cobalt porphine, [Co(I)PCO$_2$]$^{2-}$ and [Co(II)PCOOH]$^-$ are the key
intermediates.  This finding may be useful not just for electrochemical
reduction of CO$_2$, but for CO$_2$ capture from flue gas as well.  

AIMD umbrella sampling calculations show that the p$K_{\rm a}$
associated with [Co(II)PCOOH]$^-$ deprotonation is about 9.0.
This indicates that the protonation of [Co(I)PCO$_2$]$^{2-}$ is
downhill at the bicarbonate buffer experimental conditions
(pH $\sim$7).  The subsequent cleavage of the C-OH bond is also
exothermic, and the activation free energy involved is estimated to
be only 5.2~kcal/mol.  If we assume a vibrational pre-factor of
$k$=0.1~ps$^{-1}$, C-OH cleavage should occur in nanosecond timescale
at T=300~K.  Hence two key steps in the multistep reaction should
proceed readily, and it is likely that the electron transfer between
the gas diffusion electrode and the polymerized porphyrin catalyst
is the rate limiting step of the CO$_2$ to CO reduction reaction in
water.\cite{furuya,ryba,sonoyama,magd2,magd1,ramirez1,ramirez2,chilean1,chilean2}

\section*{Acknowledgement}

We thank Nicola Marzari and Heather Kulik for their input on the
self-consistent DFT+U method and Martijn Marsman for the
Wannier function VASP module.
We also thank Hank Westridge, Rick Muller, and
the principal investigators, students, and postdocs involved in this
National Institute of Nano Engineering LDRD project at Sandia,
including Nicola Spaldin, Graeme Henkelman, Jim Miller, Tiffany Hayes,
Elise Li, Zachary Pollack, and Yujiang Song.  This work was supported by the
Department of Energy under Contract DE-AC04-94AL85000.
Sandia is a multiprogram laboratory operated by Sandia
Corporation, a Lockheed Martin Company, for the U.S.~Department of Energy.  

\section*{Supporting Information Available}
Further information are provided regarding differences between the
VASP and Gaussian packages; formic acid p$K_{\rm a}$ calculation as
a benchmark of the reaction coordinate used in this paper;
the distribution Co-C-H angles in AIMD simulations;
gas phase electron affinities of reaction intermediates; and
cobalt charge state in CoPCOOH$^-$ in implicit versus explicit 
solvent environment.
This information is available free of charge via the Internet
at {\tt http://pubs.acs.org/}.

\section*{Appendix: Hysteresis in the Deprotonation AIMD Simulation}

This appendix discusses in more detail the hysteresis behavior during
p$K_a$ PMF calculations (window~3 in Fig.~\ref{fig4}).  

As mentioned in Sec.~\ref{method}, all AIMD trajectories ultimately originate
from a Monte Carlo simulation-equilibrated classical force field configuration
where a CoPCOOH is fixed in a gas phase optimized, intramolecularly
hydrogen-bonded geometry (Fig.~\ref{fig1}f).  The [Co(II)PCOOH]$^-$ is
immersed in water and equilibrated using AIMD at several stretched
values of the reaction coordinates associated with Eq.~\ref{eq5} and
Eq.~\ref{eq6} using harmonic potentials $U(R)$ (Eq.~\ref{uofr}).

During the equilibration run for the deprotonated window~2 of Fig.~\ref{fig4},
we impose $R_o=-1.12$~\AA\, on the initial [Co(II)PCOOH]$^-$
system, which turns into a [Co(I)PCO$_2$]$^{2-}$:H$_3$O$^+$
contact ion pair within 1~ps.  A maximally localized Wannier function
analysis of an AIMD snapshot confirms that, in this ion pair, Co
has spontaneously switched to the +I charge state which should be
favored for large negative $R$.  Therefore the AIMD trajectory in window~2
(and that in window~1, spawned from window~2) yields an unambiguous Co(I)
charge state.  Likewise, windows~4-7 are spawned successively from a 
window~4 trajectory equilibrated using $R_o=-0.4$~\AA, and they 
reflect a Co(II) charge state which is favorable in this $R$
range.  An additional test further confirms that the Co(II)
charge state spontaneously occurs at less negative $R$.  We start
with a [Co(I)CO$_2$]$^{2-}$, $R_o=-1.12$~\AA\, configuration in a
equilibrated window~2 trajectory and abruptly switch to $R_o=-0.4$~\AA\,
(i.e., effectively jumping from window~2 to window~4 in Fig.~\ref{fig4}).
[Co(II)PCOOH]$^-$ is recovered within 1.2~ps, consistent with the Co(II)
charge state observed in the window~4 trajectory which originally
started out as [Co(II)PCOOH]$^-$.  The change in electronic structure
in real time, represented by the changes in the spatial distribution
of the spin density, is depicted in Fig.~\ref{fig3}.  Thus, after an 
approximately 1~ps equilibration run, the final Co charge state becomes
independent of initial conditions.

Only window~3, located in the Co(I)/Co(II) transition region, exhibits
a strong dependence on initial conditions.  We have initially started from
a snapshot in window~4 and then switched $R_o=-0.4$~\AA\, to $R_o=-0.7$~\AA.
The latter value of $R_o$ does not apparently contain sufficient driving
force to rapidly alter the Co charge state, and the system remains 
[Co(II)PCOOH]$^-$ throughout the 10~ps sampling trajectory.  This yields the
dashed $W(R)$ segment in Fig.~\ref{fig4}a which exhibits a slope at the window
edge that matches poorly to the window~2, [Co(I)PCO$_2$]$^{2-}$:H$_3$O$^+$
contribution.
To make progress, we restart the window~3 simulation from a snapshot of
window~2, abruptly switch the umbrella potential from $R_o=-1.12$~\AA\,
there to $R_o$=-0.7~\AA, equilibrate for 1~ps, and collect statistics
for 10~ps.  The system remains in the Co(I) charge state throughout the
trajectory.  The corresponding $W(R)$ segment in window~3 (full curve
in Fig.~\ref{fig4}a) matches reasonably well with those in windows~2 and~4,
and is taken to be the final result.  

An ergodic AIMD simulation should in principle spontaneously sample
both Co(I) and Co(II) charge states.  Thus, the correct $W(R)$ in this
intermediate $R$ region should be a weighted average of
the full and dashed curves.  Apparently the statistical weight for 
Co(I) is much larger, so that only including Co(I) $W(R)$ information
already yields a smooth $W(R)$ curve in Fig.~\ref{fig4}.

\begin{figure}
\centerline{(a) \hbox{\epsfxsize=1.50in \epsfbox{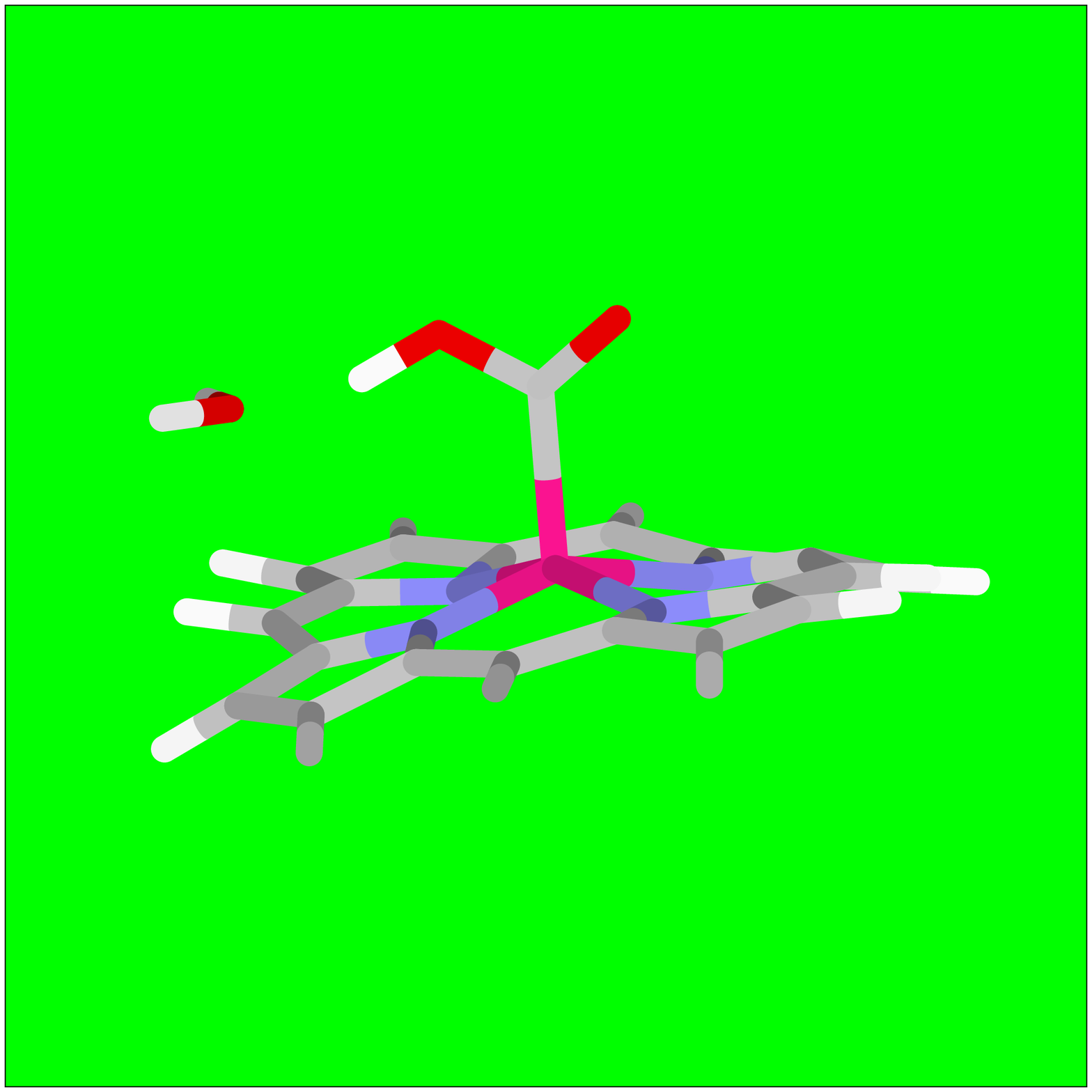}}
            \hbox{\epsfxsize=1.50in \epsfbox{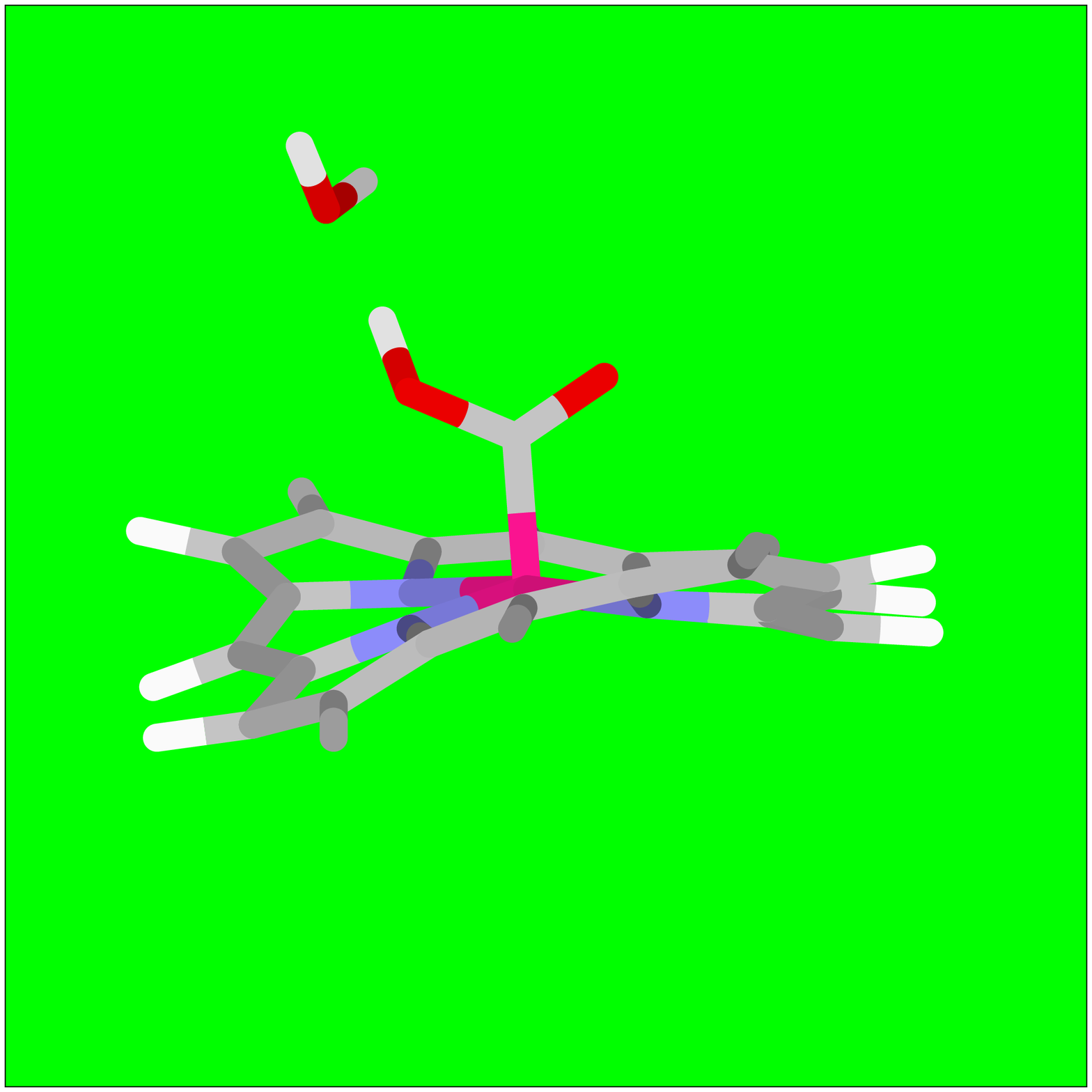}} (b)}
\caption[]
{\label{fig7} \noindent
Panels (a) \& (b) correspond to the solid and dashed line in window 3,
respectively.  The color scheme is as in Fig.~\ref{fig1}.
}
\end{figure}

A secondary, less signficant type of hysteresis associated the position of
the COOH proton also becomes apparent in Fig.~\ref{fig7}.  When the COOH group
is intact, and no H$_2$O accepts a hydrogen bond from the COOH proton,
the hydroxyl group intramolecularly hydrogen-bonds to a N~atom on the
porphine ring in a {\it cis} configuration (Fig.~\ref{fig1}f).  Enforcing
$U(R)$ with $R < 0.5$~\AA\, imposes COOH-H$_2$O hydrogen bonding that
breaks this intramolecular coupling (Fig.~\ref{fig7}a).  When $R$ is
further reduced to $R \sim -1$~\AA, the [COOH]$^-$
proton is detached from the hydroxyl oxygen, landing on the hydrogen-bonding
accepting H$_2$O molecule.  This newly formed H$_3$O$^+$ spontaneously migrates
away from the hydrophobic porphine ring and coordinates to what is now
a CO$_2^{\delta-}$ group in the axial ({\it trans}) position.  Now the excess
proton sticks out of the porphine plane.  When we increase $R_o$ to reprotonate
the COOH group, the system remains in this isomeric form (Fig.~\ref{fig7}b).
Isomerization between the two may entail a free energy barrier of several
kcal/mol even for a stretched O-H, and does not occur on AIMD timescales
in window~3.  The lack of isomerization should have no significant
effect on the $W(R)$ of window~3, while the gas phase Fig.~\ref{fig1}f
[Co(II)PCOOH]$^-$ intramolecular hydrogen bonded configuration is
stabilized over the Fig.~\ref{fig1}e isomer by about 4.93~kcal/mol, that
hydrogen bond is already broken when the carboxylate proton
donates a hydrogen bond to a water molecular (Fig.~\ref{fig7}a).  This
secondary hysteresis may be avoided altogether using an artificial
deprotonation coordinate,\cite{sprik_new_proton} although it is not obvious
the electronic hysteresis will be avoided as well.

\begin{references}
\bibitem{note98}
Unlike Paper~I,\cite{paper1} Eq.~\ref{eq3} is written without a second
proton; it is trivially related to the CoPCOOH + H$^+$ $\rightarrow$
CoPCO + H$_2$O reaction studied in Paper I via the p$K_w$ of water.
\bibitem{note9}
As DFT+U calculations can be difficult to converge, we first use
the robust Davison method in VASP (``IALG=38'') to compute the
Kohn-Sham wavefunctions of the first ionic step and then apply these
as starting points to propagate subsequent molecular dynamics steps
using the faster conjugate gradient (``IALG=48'') algorithm.
Gas phase DFT+U calculations do not converge for some of the
doubly negatively charged species.
\bibitem{note99}
The same consideration should be used for the deprotonation $W(R)$.
However, the reaction coordinate is more complex there, and 
$R_{\rm O-H}$ only varies by about 0.5~\AA\, in that case; therefore
the rotational prefactor $4\pi R_{\rm O-H}^2$ varies less signficantly.
We have not frozen any degrees of freedom in that calculation but have
included a $4\pi R_{\rm O-H}^2$ factor in the deprotonation free energy.
\bibitem{note_cop}
Maximally localized Wannier functional analyses of CoP$^-$ and CoP$^{2-}$
in water reveal that both species exhibit 8 filled Co $3d$ spin-orbitals,
consistent with a Co(I) charge state.
\end{references}
\end{document}